\documentclass[12pt]{article}
\pdfoutput=1
\usepackage{graphicx}
\usepackage{amsmath}
\usepackage{amssymb}
\usepackage{amscd}
\usepackage{amsthm}
\usepackage{amssymb}
\usepackage{fullpage}
\usepackage{setspace}
\usepackage[hang,small]{caption}
\usepackage{authblk}
\usepackage{subfigure}
\def \be {\begin{equation}}
\def \ee {\end{equation}}
\def \bea {\begin{eqnarray}}
\def \eea {\end{eqnarray}}
\numberwithin{equation}{section}
\title{\textbf{Causality and Hyperbolicity of \\Lovelock Theories}}

\author{Harvey S. Reall$^a$, Norihiro Tanahashi$^{b,a}$, Benson Way$^a$}
\affil{ \textit{$^a$DAMTP, Centre for Mathematical Sciences, University of Cambridge, Wilberforce Road, Cambridge CB3 0WA, UK}}
\affil{ 
\textit{$^b$Kavli Institute for the Physics and Mathematics of the Universe,}\\
\textit{Todai Institutes for Advanced Study, University of Tokyo (WPI),}\\
\textit{5-1-5 Kashiwanoha, Kashiwa, Chiba 277-8583, Japan}}

\date{}
\begin{document}
\maketitle
\begin{abstract}
In Lovelock theories, gravity can travel faster or slower than light. The causal structure is determined by the characteristic hypersurfaces. We generalise a recent result of Izumi to prove that any Killing horizon is a characteristic hypersurface for all gravitational degrees of freedom of a Lovelock theory. Hence gravitational signals cannot escape from the region inside such a horizon. We investigate the hyperbolicity of Lovelock theories by determining the characteristic hypersurfaces for various backgrounds. First we consider Ricci flat type N spacetimes. We show that characteristic hypersurfaces are generically all non-null and that Lovelock theories are hyperbolic in any such spacetime. Next we consider static, maximally symmetric black hole solutions of Lovelock theories. Again, characteristic surfaces are generically non-null. For some small black holes, hyperbolicity is violated near the horizon. This implies that the stability of such black holes is not a well-posed problem.  
\end{abstract}
\onehalfspacing
\section{Introduction}

The Einstein equation relates the curvature of spacetime to the energy-momentum tensor of matter:
\be
 G_{ab} + \Lambda g_{ab} = 8\pi T_{ab}
\;.\ee
The form of the LHS of this equation is dictated by Lovelock's theorem \cite{lovelock}. This states that in four dimensions, the most general symmetric, divergence-free, second rank tensor that is a function of only the metric and its first and second derivatives, is a linear combination of the Einstein tensor and a cosmological constant term. 

In $d>4$ dimensions, this result is not valid and Lovelock showed that additional terms can appear in the LHS above. Theories with these additional terms are referred to as Lovelock theories. The Einstein equation is obtained only if one adds the additional criterion that the equation of motion should be linear in second derivatives of the metric, i.e.\ that the equation of motion is quasilinear. 

General Lovelock theories are not quasilinear, which makes them rather exotic. Basic properties of these theories are unclear. For example, when $\Lambda=0$, is Minkowski spacetime (nonlinearly) stable in Lovelock theories? Do such theories admit a positive energy theorem? Is the initial value problem well-posed?

The basic causal properties of a system of PDEs are governed by its characteristic hypersurfaces. For the Einstein equation, a hypersurface is characteristic if, and only if, it is null: ``gravity travels at the speed of light". Characteristics hypersurfaces of Lovelock theories were investigated in Refs.~\cite{aragone,CB}. In particular, Ref.~\cite{CB} showed that such surfaces are generically non-null. Gravity can propagate faster (or slower) than light.\footnote{Superluminal propagation of gravitons has played an important role in discussions of asymptotically anti-de Sitter (AdS) black hole solutions of Einstein-Gauss-Bonnet theory \cite{brigante,brigante2,hofman}. Shock-wave solutions have also been considered \cite{hofman}. These studies investigated when bulk superluminal propagation would lead to physically unacceptable superluminal propagation in a putative dual CFT. It was found that this requirement imposes constraints on the Gauss-Bonnet coupling constant. Similar results have been obtained for other Lovelock theories, see e.g.~\cite{Camanho:2009hu,Camanho:2010ru}.}

This raises the possibility that gravitational signals could escape from the interior of a Lovelock black hole. Very recently, Izumi has proved a result suggesting that this does not happen for stationary black holes in Einstein-Gauss-Bonnet theory (a Lovelock theory) \cite{izumi}. He proved that a Killing horizon is a characteristic hypersurface for all gravitational degrees of freedom. If one assumes that the event horizon of a stationary black hole is a Killing horizon then it follows that gravitational signals cannot escape from the black hole interior. In this paper, we will generalize this result to an arbitrary Lovelock theory.

The possibility of spacelike characteristic surfaces entails problems for the initial value problem in Lovelock theories. In order to evolve in time, the initial data should be non-characteristic. For such data, a unique solution to the equations of motion may exist locally (although this has been proved only for the special case of analytic initial data \cite{CB}). But the non-characteristic condition might be violated after a finite time, in which case one cannot evolve further. This issue does not arise for the Einstein equations, for which any spacelike surface is non-characteristic. However, it can happen in Lovelock theories e.g.\ this behaviour has been observed for cosmological solutions \cite{Deruelle:1989fj}. If this happens, one would have to investigate whether it is possible to evolve further by choosing a different time slicing. 

A closely related issue concerns the canonical formulation of Lovelock theories. The lack of quasilinearity implies that the canonical momentum $\pi^{ij}$ depends on the extrinsic curvature $K_{ij}$ (of a constant time hypersurface) in a nonlinear (polynomial) manner \cite{teitelboim}. This relation is generically non-invertible. However, using the inverse function theorem, one can choose a branch such that $K_{ij}$ depends smoothly on $\pi^{ij}$, provided that the constant time surfaces are non-characteristic. If such a surface becomes characteristic then even this local invertibility fails.

Another basic question about Lovelock theories concerns their hyperbolicity. If the spacetime curvature is small (with respect to scales defined by the coupling constants of the theory) then these theories will be hyperbolic.\footnote{
We will not consider Lovelock theories with vanishing coefficient of the Einstein tensor in the equation of motion.}
However, when the curvature becomes large then hyperbolicity may fail. One aim of this paper is to investigate whether this happens. To do this, we will investigate the characteristic surfaces of certain solutions. This also seems of interest in its own right since there are not many non-trivial examples for which the characteristic hypersurfaces have been determined.  

We start by considering spacetimes that are Ricci flat with a Weyl tensor of type N in the classification of Ref.~\cite{cmpp}. We note that any such spacetime is a solution of Lovelock theory (with $\Lambda=0$) (slightly extending a result of \cite{pravda}). In this case, we find that one can define $d(d-3)/2$ Lorentzian metrics such that the characteristic surfaces are surfaces null with respect to any of these metrics.  With respect to the physical metric, the characteristic surfaces are generically non-null. Note that $d(d-3)/2$ is the number of degrees of freedom of the graviton. The null cones of these metrics form a nested set. This result implies that Lovelock theory is hyperbolic in any such background, no matter how large the curvature.

Next we consider static, spherically (or planar) symmetric black hole solutions of these theories \cite{boulware,wheeler,cai,cai2}. One can determine characteristics from the equations of motion of linearized perturbations of a background. For these geometries, such perturbations can be decomposed into scalar, vector and tensor types. Each of these satisfies a certain ``master equation" \cite{tensorperts,tensorperts2,scalarperts,Konoplya:2008ix,takahashi,takahashi2,Takahashi:2010ye,Takahashi:2010gz}. From this one can infer that there exist ``effective metrics" $G^{ab}_S$, $G^{ab}_V$ and $G^{ab}_T$ such that characteristic hypersurfaces are null hypersurfaces with respect to one of these metrics. It turns out that, for certain small Lovelock black holes, one of these metrics may change signature near, but outside, the horizon of the black hole. This is equivalent to the observation that some perturbations become ``ghost-like" near the black hole \cite{Takahashi:2010gz}.

When this occurs, the linearized equation admits solutions which grow exponentially with time so the change of sign has been interpreted as indicating an instability of the black hole. But actually it is much worse. Asking about stability presupposes that the equations admit a well-posed initial value formalism for initial data in a neighbourhood of the black hole initial data. But if the equations are not hyperbolic then it seems unlikely that the initial value problem for small perturbations is well-posed. We will argue that (i) solutions of the linearized equations of motion do not depend continuously on their initial data; (ii) for a generic smooth initial perturbation there does not exist any corresponding solution of the linearized equations of motion. Hence the initial value problem is ill-posed: time evolution does not make sense here. We will discuss the implications of this at the end of this paper.

This paper is organized as follows. In section \ref{sec:background} we review Lovelock theories and the definition of characteristic hypersurfaces. Section \ref{sec:lovelockcharacteristics} discusses the characteristic hypersurfaces of Lovelock theories, proves that a Killing horizon is characteristic in such theories, and reviews the definition of hyperbolicity. Section \ref{sec:typeN} studies Ricci flat type N spacetimes. Section \ref{sec:bh} studies static black hole solutions. Finally, section \ref{sec:discuss} contains further discussion. 

\section{Background}

\label{sec:background}

\subsection{Lovelock theories}

The most general symmetric tensor that (in a coordinate chart) (i) depends only on the metric and its first and second derivatives and (ii) is conserved is \cite{lovelock}\footnote{
Conventions: Latin indices are abstract indices, Greek indices refer to a particular basis.
}
\be
 A^a{}_b =  \Lambda \delta^a_b+ k_1 G^a{}_b  + B^a{}_b\;,
\ee
where
\be
 B^a{}_b = \sum_{p \ge 2} k_p \delta^{a c_1 \ldots c_{2p}}_{b d_1 \ldots d_{2p}} R_{c_1 c_2}{}^{d_1 d_2} \ldots R_{c_{2p-1} c_{2p}}{}^{d_{2p-1} d_{2p}}\;,
\ee
and $k_p$ are constants. The antisymmetry ensures that the sum is finite ($2p+1 \le d$ in $d$ dimensions). Note that
\be
\label{bianchiB}
 \nabla^b B_{ab} = 0\;.
\ee
The generalisation of the Einstein equation is
\be
 A_{ab} = 8 \pi T_{ab}\;.
\ee
We will assume that $k_1 > 0$ and choose units so that $k_1=1$.  The Lagrangian density for Lovelock theory is \cite{lovelock}
\be
 {\cal L } =\sqrt{-g} \left( R -2 \Lambda \right) - \sqrt{-g} \sum_{p \ge 2} 2 k_p \delta^{c_1 \ldots c_{2p}}_{d_1 \ldots d_{2p}} R_{c_1 c_2}{}^{d_1 d_2} \ldots R_{c_{2p-1} c_{2p}}{}^{d_{2p-1} d_{2p}} \;.
\ee
If we retain only the $p=2$ term above then we have Einstein-Gauss-Bonnet theory. 

\subsection{Characteristics}

\label{sec:characteristics}

In this section we will review the definition and basic theory of characteristic hypersurfaces. Consider a field theory in which the unknown fields form a column vector $g_I$ with equation of motion
\be
 E_I \left( g,\partial g, \partial^2 g \right)=0
\;.\ee
(In a Lovelock theory $g_I$ will stand for the metric $g_{\mu\nu}$.) The theory is quasilinear if $E_I$ is linear in $\partial^2 g_J$. We will not assume this. However, Lovelock theories have the special property\footnote{This was noticed in \cite{aragone,CB} for coordinates adapted to a spacelike surface but it holds generally.}  that, in any coordinate chart $x^\mu$, the equations of motion depend linearly on $\partial_0^2 g_{\mu\nu}$. So we will assume that $E_I$ has this property. Hence in any chart the equation of motion takes the form
\be
\label{eqofmotion}
 \frac{\partial^2 E_I}{\partial (\partial_0^2 g_J)} \partial_0^2 g_J + \cdots = 0
\;.\ee
where the ellipsis denotes terms involving fewer than 2 derivatives with respect to $x^0$ and the coefficient of $\partial_0^2 g_J$ does not depend on $\partial_0^2 g_J$. 

Now consider a hypersurface $\Sigma$ and introduce adapted coordinates $(x^0,x^i)$ so that $\Sigma$ has equation $x^0=0$. Assume that $g_I$ and $\partial_\mu g_I$ are known on $\Sigma$. By acting with $\partial_i$ we then also know $\partial_i \partial_\mu g_I$ on $\Sigma$. The only second derivatives that we don't know are $\partial_0^2 g_I$. These are uniquely determined by the equation of motion (\ref{eqofmotion}) if, and only if, the matrix
\be
\label{principaladapted}
  \frac{\partial E_I}{\partial (\partial_0^2 g_J)}
\ee
is invertible. If this is the case then $\Sigma$ is said to be non-characteristic. If the matrix is {\it not} invertible anywhere on $\Sigma$ then $\Sigma$ is characteristic. Let $\xi = dx^0$ be the normal to $\Sigma$. We can write the above matrix covariantly as
\be
 P(x,\xi)_I{}^J = \frac{\partial E_I}{\partial (\partial_\mu \partial_\nu g_J)} \xi_\mu \xi_\nu
\;.\ee
This is called the {\it principal symbol} of the equation. The {\it characteristic polynomial} is
\be
 Q(x,\xi) = \det P(x,\xi)
\;.\ee
A surface with normal $\xi$ is characteristic if, and only if, $Q(x,\xi)=0$ vanishes everywhere on the surface. $Q$ is a homogeneous polynomial in $\xi$. The equation $Q=0$ at a point $p$ defines the {\it normal cone} at $p$. 

A surface $\phi(x)={\rm constant}$ is characteristic if $Q(x,d\phi)=0$. This is a first order PDE for $\phi$. The theory of first order PDEs implies that such surfaces are generated by bicharacteristic curves $(x^\mu(t),\xi_\nu(t))$ defined by \cite{courant}
\be
 \dot{x}^\mu = \frac{\partial Q}{\partial \xi_\mu}\;, \qquad \dot{\xi}_\mu = -\frac{\partial Q}{\partial x^\mu}\;,
 \ee
with the initial values of $\xi_\mu$ chosen so that $Q=0$ (this is preserved along the curves). The {\it ray cone} at $p$ is defined as the set of vectors of the form $\partial Q/\partial \xi_\mu$ for $\xi$ obeying $Q=0$.

If $Q$ factorizes (for arbitrary $\xi$) into a product of polynomials of lower degree
\be
\label{Qfact}
 Q = Q_1^{p_1} Q_2^{p_2} \ldots
\;,\ee
then we must use $Q_i$ instead of $Q$ in defining bicharacteristics. This happens in GR: writing the Einstein equation in harmonic coordinates gives $Q = Q_1^{d(d+1)/2}$ where $Q_1 = -(1/2) g^{ab} \xi_a \xi_b$. In this case the curves $x^\mu(t)$ are the null geodesics of $g_{ab}$ and a surface is characteristic if, and only if, it is null.

To understand the role of characteristic hypersurfaces as wavefronts, consider a solution which is smooth everywhere except across a hypersurface $\Sigma$ on which the solution is $C^1$ but $\partial^2 g_I$ is discontinuous. In this case, the equation of motion cannot determine uniquely $\partial^2 g_I$ on $\Sigma$. Hence $\Sigma$ must be a characteristic surface. So discontinuities in $\partial^2 g_I$ must propagate along characteristic hypersurfaces. 

By taking derivatives of the equation of motion one easily sees that discontinuities in $\partial^k g_I$, $k \ge 3$ also propagate along characteristic hypersurfaces, i.e., if a solution is smooth on either side of $\Sigma$ and $C^{k-1}$ on $\Sigma$ with a discontinuity in $\partial^k g_I$ on $\Sigma$ then $\Sigma$ must be characteristic. 

\section{Characteristics of Lovelock theories}

\label{sec:lovelockcharacteristics}

\subsection{General properties}

The characteristics of Lovelock theories were discussed by Aragone \cite{aragone} and in more detail by Choquet-Bruhat \cite{CB}. These references studied characteristics by imposing various conditions on the coordinates (e.g.\ an ADM type decomposition or harmonic coordinates). Recently, Ref. \cite{izumi} studied characteristics of Einstein-Gauss-Bonnet theory using a first order approach. We will use a second order, gauge-invariant, approach.
 
Consider Lovelock theory in vacuum $T_{ab}=0$, so $A_{ab}=0$ or equivalently
\be
\label{eofm}
 E_{ab} \equiv R_{ab} - \frac{2\Lambda}{d-2} g_{ab} + B_{ab} - \frac{1}{d-2} B^c{}_c g_{ab}  = 0\;.
\ee
In a coordinate chart $x^\mu$, the principal symbol is defined for a 1-form $\xi_a$ by
\be
 P(x,\xi)_{\mu\nu}{}^{\rho\sigma} = \frac{\delta E_{\mu\nu} }{\delta (\partial_\alpha \partial_\beta g_{\rho\sigma})} \xi_\alpha \xi_\beta
\;,\ee 
which we view as a matrix mapping a symmetric tensor $t_{\rho\sigma}$ to a symmetric tensor $P_{\mu\nu}{}^{\rho \sigma} t_{\rho \sigma}$. To calculate $P$ we vary $g_{\mu\nu} \rightarrow g_{\mu\nu} + \delta g_{\mu\nu}$ and use
\be
 \delta B^\mu{}_\nu = -\sum_{p \ge 2} 2p k_p \delta^{\mu \rho_1 \ldots \rho_{2p}}_{\nu \sigma_1 \ldots \sigma_{2p}}( \partial_{\rho_1} \partial^{\sigma_1} \delta g_{\rho_2}{}^{\sigma_2}) R_{\rho_3 \rho_4}{}^{\sigma_3 \sigma_4} \dots R_{\rho_{2p-1} \rho_{2p}}{}^{\sigma_{2p-1} \sigma_{2p}} + \cdots
\;,\ee
where the ellipsis denotes terms that don't involve second derivatives of $\delta g_{\mu\nu}$.  This gives
\be
 (P \cdot t)^a{}_b =  (P_{GR}  \cdot t)^a{}_b+ ({\cal R} \cdot t)^a{}_b
\;, \ee
where the term coming from the Ricci tensor in (\ref{eofm}) is the same as for GR:
\be
 (P_{GR} \cdot t)_{ab} = -\frac{1}{2} \xi^2 t_{ab} + \xi^c \xi_{(a} t_{b)c} - \frac{1}{2} \xi_a \xi_b t^c{}_c\;,
\ee
and the other piece is
\bea
  ({\cal R} \cdot t)^a{}_b &=&
 -\sum_{p \ge 2} 2p k_p \delta^{a c_1 \ldots c_{2p}}_{b d_1 \ldots d_{2p}} \xi_{c_1} \xi^{d_1} t_{c_2}{}^{d_2} R_{c_3 c_4}{}^{d_3 d_4} \dots R_{c_{2p-1} c_{2p}}{}^{d_{2p-1} d_{2p}} \nonumber \\
&+& \frac{1}{d-2} \delta^a_b \sum_{p \ge 2} 2p k_p \delta^{e c_1 \ldots c_{2p}}_{e d_1 \ldots d_{2p}} \xi_{c_1} \xi^{d_1} t_{c_2}{}^{d_2} R_{c_3 c_4}{}^{d_3 d_4} \dots R_{c_{2p-1} c_{2p}}{}^{d_{2p-1} d_{2p}}
\;,\eea
and we are now using abstract indices because these expressions are valid in any basis. 

$P_{GR}(x,\xi)$ and ${\cal R}(x,\xi)$ are linear operators which map symmetric tensors to symmetric tensors. Define an inner product between symmetric tensors 
\be
\label{innerproduct}
 (t , t') =G^{abcd} t_{ab} t_{cd}' \equiv  t^{ab} t_{ab}' -\frac{1}{2} t^a{}_a t^{'b}{}_b
\;,\ee
where
\be
 G^{abcd} = \frac{1}{2} \left(g^{ac} g^{bd} + g^{ad} g^{bc} - g^{ab} g^{cd} \right)
\;.\ee
$P_{GR}$ and ${\cal R}$ are symmetric with respect to this inner product, and hence so is $P$:
\be
(P \cdot t,t') = (t,P \cdot t')
\;.\ee
By decomposing $t$ into a traceless part $\hat{t}$ and a trace, and working in an orthonormal basis $\{ e_0,e_i \}$, one finds that the inner product has signature $(-,-,\ldots ,-,+,+,\ldots +)$ where there are $d$ negative eigenvalues and $d(d-1)/2$ positive eigenvalues. The negative eigenvalues are associated to the trace and to the components $\hat{t}_{0i}$, $i=1, \ldots, d-1$. 

The principal symbol $P(x,\xi)$ is always degenerate. This is because the equations have a gauge symmetry arising from diffeomorphisms, which implies that $P \cdot t$ is invariant under
\be
 t_{ab} \rightarrow t_{ab} + \xi_{(a} X_{b)}
\ee
for any $X_b$. We will deal with this by working with equivalence classes of symmetric tensors. We will say that $t'_{ab} \sim t_{ab}$ if $t'_{ab} = t_{ab} + \xi_{(a} X_{b)}$ for some $X_b$. This defines an equivalence relation on symmetric tensors. Let $V_{\rm physical}(\xi)$ be the vector space of equivalence classes with respect to this equivalence relation. Then gauge symmetry implies that $P_{GR}$ and ${\cal R}$ are well-defined on $V_{\rm physical}$. 

We will say that a symmetric tensor $t_{ab}$ is {\it transverse} with respect to $\xi_a$ if it obeys
\be
\label{transverse}
 \xi^b t_{ab} - \frac{1}{2} \xi_a t^c{}_c =0\;.
\ee
Let $V_{\rm transverse}(\xi)$ denote the vector space of transverse symmetric tensors. The Bianchi identities imply that $P_{GR} \cdot t$ and ${\cal R} \cdot t$ lie in $V_{\rm transverse}$ for any $t_{ab}$. Hence we can regard $P_{GR}$, ${\cal R}$ and $P$ as maps from $V_{\rm physical}$ to $V_{\rm transverse}$:
\be
 P_{GR}(x,\xi), {\cal R}(x,\xi), P(x,\xi) : ~ V_{\rm physical} \rightarrow V_{\rm transverse}
 \;.\ee
Note that $V_{\rm physical}$ and $V_{\rm transverse}$ both have dimension $d(d-1)/2$. If we pick bases for these spaces then we can define the characteristic polynomial as
\be
 Q(x,\xi) = \det P(x,\xi)\;.
\ee
This is a homogeneous polynomial in $\xi_a$ of degree $d(d-1)$. For generic $\xi_a$ it will be non-zero. A hypersurface with normal $\xi_a$ is characteristic if and only if $Q(x,\xi)=0$. 

\subsection{Characteristics of GR}

\label{sec:GR}

Let's see how this works for GR, with principal symbol $P=P_{GR}$. If $Q(x,\xi)=0$ then $P_{GR}(x,\xi)$ is degenerate.  Hence there exists a non-zero element of $V_{\rm physical}$ that is annihilated by $P_{GR}(x,\xi)$. Let $t_{ab}$ be an element of this equivalence class, so $P_{GR}(x,\xi )\cdot t =0$:
\be
-\frac{1}{2} \xi^2 t_{ab} + \xi^c \xi_{(a} t_{b)c} - \frac{1}{2} \xi_a \xi_b t^c{}_c =0\;.
\ee
 If $\xi^2 \ne 0$ then this equation implies that $t_{ab} = \xi_{(a} X_{b)}$ for some $X_b$, which is a contradiction because this is pure gauge and hence corresponds to the zero element of $V_{\rm physical}$. Hence any characteristic direction must be null: $\xi^2=0$. The above equation then implies (\ref{transverse}) i.e.\ $t_{ab}$ is transverse. Note that (\ref{transverse}) is gauge invariant if $\xi^2=0$ so this condition defines a subspace $V_{\rm physical transverse}$ of $V_{\rm physical}$. This subspace has dimensions $d(d-1)/2-d = d(d-3)/2$. Hence we have shown that, for GR, $\xi_a$ is characteristic if and only if $\xi^2=0$ and, for such $\xi_a$, the kernel of $P_{GR}(x,\xi)$ is the subspace $V_{\rm physical transverse}$ with dimension $d(d-3)/2$, which is the number of physical degrees of freedom of the graviton.

\subsection{Characteristics of Lovelock theories}
\label{sec:lovelockcharacteristic}

Now consider a Lovelock theory with $P=P_{GR} + {\cal R}$. Let $\xi_a$ satisfy $Q(x,\xi)=0$. As above, for this $\xi_a$, there exists a non-zero element of $V_{\rm physical}$ that is annihilated by $P(x,\xi)$. Let $t_{ab}$ be an element of this equivalence class. 

Consider first the case in which $\xi_a$ is non-null: $\xi^2 \ne 0$. In this case we can decompose any $t_{ab}$ uniquely into a transverse part and a gauge part:
\be
 t_{ab} = \hat{t}_{ab} + \xi_{(a} X_{b)} \;,
\ee
where $\hat{t}_{ab}$ satisfies (\ref{transverse}) and is non-zero (since $t_{ab}$ is not pure gauge). We then have $0=P(x,\xi) \cdot t = P(x,\xi) \cdot \hat{t}$, which reduces to the eigenvalue equation 
\be
\label{evaleq}
 {\cal R}(x,\xi) \cdot \hat{t} = \frac{1}{2} \xi^2 \hat{t}\;.
\ee
Hence non-null $\xi_a$ is characteristic if and only if ${\cal R}(x,\xi)$ admits an eigenvector $\hat{t}_{ab} \in V_{\rm transverse}$ with eigenvalue $(1/2) \xi^2$. Here we regard ${\cal R}(x,\xi)$ can be regarded as a map from $V_{\rm transverse}$ to itself so the problem of determining non-null characteristics is the eigenvalue problem for this map.

Now consider the case in which $\xi_a$ is characteristic and null. Introduce a null basis $\{e_0,e_1,e_i \}$ $(i = 2, \dots, d-1$) where $e_0^a = \xi^a$, $e_1$ is null with $e_0 \cdot e_1 = 1$, and $e_i$ are spacelike, orthonormal and orthogonal to $e_0$ and $e_1$. In other words the only non-vanishing inner products between the basis vectors are
\be
 e_0 \cdot e_1 = 1\;, \qquad e_i \cdot e_j = \delta_{ij}\;.
\ee
The non-trivial components of the equation $P(x,\xi) \cdot t = 0$ are
\begin{subequations}\label{cpt}
\be
\label{01cpt}
 \frac{1}{2} t_{00} +\left(  {\cal R} \cdot t \right)_{01} = 0\;,
\ee
\be
\label{ijcpt}
 \left( {\cal R} \cdot t \right)_{ij} = 0\;,
\ee
\be
\label{1icpt}
 \frac{1}{2} t_{0i} + \left( {\cal R} \cdot t \right)_{1i} = 0\;,
\ee
\be
\label{11cpt}
-\frac{1}{2} t_{ii} 
+ \left( {\cal R} \cdot t \right)_{11} = 0\;,
\ee
\end{subequations}
and note that ${\cal R} \cdot t \in V_{\rm transverse}$ is equivalent to 
\be
({\cal R} \cdot t)_{00} = ({\cal R} \cdot t )_{0i} = ({\cal R} \cdot t)_{ii} = 0\;,
\ee
so the LHS of (\ref{ijcpt}) is traceless. The components $t_{1\mu}$ are ``pure gauge" and hence do not appear in the above equations. The other $d(d-1)/2$ components are gauge invariant and governed by the $d(d-1)/2$ equations 
\eqref{cpt}. 
Since the number of equations equals the number of unknowns, we would not expect any non-zero solution to these equations at a generic point of a generic spacetime. So generically, there do not exist null characteristic directions. However, as we will see now, there are special circumstances under which one can have null characteristic directions, corresponding to a non-zero solution of the above equations.

\subsection{Application: Killing horizons are characteristic}

Gravitational signals can propagate faster than light in Lovelock theories. This raises the question of whether gravitational signals can escape from the interior of a black hole in such theories. Recently, Izumi has argued that a Killing horizon is a characteristic hypersurface in Einstein-Gauss-Bonnet theory \cite{izumi}. If one assumes that the event horizon of a stationary black hole is a Killing horizon\footnote{
This assumption is justified for a {\it static} black hole subject to some reasonable global assumptions \cite{hawking}.} then this indicates that gravitational signals cannot escape from the interior of a stationary black hole in Einstein-Gauss-Bonnet theory. In this section, we will generalize Izumi's result to any Lovelock theory. More precisely we will prove the following

\medskip

\noindent {\bf Proposition.} Consider a solution of a Lovelock theory containing a Killing horizon ${\cal H}$ with associated Killing vector field $\xi^a$. Then the null hypersurface ${\cal H}$, with normal $\xi_a$, is a characteristic hypersurface. On this hypersurface, the kernel of $P(x,\xi)$ has dimension $d(d-3)/2$, so ${\cal H}$ is ``characteristic for all gravitational degrees of freedom". 

\medskip

\noindent {\it Proof.} Introduce a null basis with $e_0 = \xi$ as in the previous subsection. We need to show that there exist $d(d-3)/2$ independent solutions of the system \eqref{cpt}. In this basis, we have
\bea
\label{Rkilling}
({\cal R} \cdot t)^\mu{}_\nu &=& - \sum_{p\ge 2} 2p k_p \delta^{\mu 1 \rho_2 \ldots \rho_{2p}}_{\nu 0 \sigma_2 \ldots \sigma_{2p}} t_{\rho_2}{}^{\sigma_2} R_{\rho_3 \rho_4}{}^{\sigma_3 \sigma_4} \ldots R_{\rho_{2p-1} \rho_{2p}}{}^{\sigma_{2p-1} \sigma_{2p}} \nonumber \\ &+& \frac{1}{d-2} \delta^\mu_\nu \sum_{p\ge 2} 2p k_p \delta^{i 1 \rho_2 \ldots \rho_{2p}}_{i 0 \sigma_2 \ldots \sigma_{2p}} t_{\rho_2}{}^{\sigma_2} R_{\rho_3 \rho_4}{}^{\sigma_3 \sigma_4} \ldots R_{\rho_{2p-1} \rho_{2p}}{}^{\sigma_{2p-1} \sigma_{2p}}\;.
\eea
Since ${\cal H}$ is a Killing horizon, its generators are free of expansion, rotation and shear \cite{wald}. This implies that the Riemann tensor obeys\footnote{This follows e.g.\ from equations NP1, NP2, NP3 of \cite{GHP}.}
\be
\label{killingcurv}
R_{0i0j} = R_{0ijk}=0\;.
\ee
Assume that 
\be
\label{tkilling}
t_{00} = t_{0i} = 0\;.
\ee
Recall that $t_{1 \nu}$ is pure gauge and so does not contribute to (\ref{Rkilling}). Let $\mu \ne 0$. Then for the Kronecker deltas in (\ref{Rkilling}) to be non-zero we need one of the upper $\rho$ indices to be $0$. But then the conditions (\ref{killingcurv}) and (\ref{tkilling}) imply that this expression must vanish. Since $({\cal R} \cdot t)_{\mu\nu}$ is symmetric, it follows that the only non-vanishing component is the $\mu=\nu=1$ component. So (\ref{01cpt}), (\ref{ijcpt}) and (\ref{1icpt}) are all satisfied and (\ref{11cpt}) is the only non-trivial equation. So any $t_{\mu\nu}$ satisfying the $d-1$ conditions (\ref{tkilling}) and the single condition (\ref{11cpt}) gives a solution. 
These are
$d$ conditions in total. $V_{\rm physical}$ has dimension $d(d-1)/2$. Hence the size of the kernel is $d(d-1)/2 - d = d(d-3)/2$ as claimed.\footnote{We have not excluded the possibility that the kernel is bigger than this, i.e., that there may exist solutions of the system (\ref{01cpt}) to (\ref{11cpt}) that do not satisfy (\ref{tkilling}). But the above analysis shows that (\ref{01cpt}), (\ref{ijcpt}) and (\ref{1icpt}) depend only on $t_{00}$ and $t_{0i}$ hence they form an overdetermined system for these quantities. Therefore it seems likely that all solutions must satisfy (\ref{tkilling}).}

\subsection{Hyperbolicity}

We can now discuss hyperbolicity. The idea that we want to capture is that, given a suitable ``initial" hypersurface $\Sigma$, for any $(d-2)$-dimensional surface $S$ within $\Sigma$ there are $d(d-3)$ physical characteristic hypersurfaces through $S$. These correspond roughly to ``ingoing" and ``outgoing" wavefronts for each of the $d(d-3)/2$ physical polarizations of the graviton. In the PDE literature, such a surface $\Sigma$ is referred to as spacelike \cite{courant} but, since we already have a notion of spacelike arising from the metric, we will refer to it as ``Lovelock-spacelike". 

More precisely, consider a basis of 1-forms $\{f^{(\mu)}_a \}$ for the cotangent space at $p$ such that $f^{(0)}$ is normal to $\Sigma$. We can expand $\xi = \xi_\mu f^{(\mu)}=\xi_0 f^{(0)}+ \xi_i f^{(i)}$. We will say that $\Sigma$ is ``Lovelock-spacelike" at $p$ if, for every $\xi_i \ne 0$ the equation $Q(x^i,\xi_0,\xi_i)=0$ has exactly $d(d-3)$ distinct real roots $\xi_0$. If a Lovelock-spacelike hypersurface exists through a point $p$ then the theory is hyperbolic at $p$. Note that it is possible that the theory could be hyperbolic in some region of spacetime but non-hyperbolic in another region. We will see an example of this below.

This definition can be extended to permit degeneracy of the roots $\xi_0$. If $\xi_0$ has degeneracy $k$ then we require that there should be $k$ modes propagating along the corresponding characteristic surface. We do this by 
requiring that there exist $k$ linearly independent $t_{ab} \in V_{\rm physical}$ that belong to the kernel of $P(x,\xi)$. 

For some spacetimes (e.g.\ those with appropriate symmetries) the characteristic polynomial factorizes into a product of quadratic factors:
\be
\label{Qfact2}
 Q(x,\xi) = \left( G_1^{ab} (x) \xi_a \xi_b \right)^{p_1}  \left( G_2^{ab} (x) \xi_a \xi_b \right)^{p_2} \ldots
\ee
for certain symmetric tensors $G^{ab}_I$, which can be regarded as (inverse) ``effective" metric tensors. In this case, the normal cone is the product of the null cones of each of these metrics. A hypersurface is characteristic if and only if it is null with respect to one of these metrics, and the bicharacteristic curves are the null geodesics of these metrics. In a generic spacetime, there is no reason for such factorization to occur and the characteristic cone will not be a product of null cones \cite{CB}. Even when factorisation does occur, the tensors $G^{ab}_I$ might be degenerate or have non-Lorentzian signature, in which case the theory would not by hyperbolic in such a background.

In GR, the equation of motion is hyperbolic everywhere. However, in a Lovelock theory hyperbolicity may fail somewhere in the spacetime. For example, if $Q$ takes the form (\ref{Qfact2}) then hyperbolicity will fail if any of the tensors $G^{ab}_I$ fails to be a Lorentzian metric. We will show 
that this happens near the horizon of certain black hole solutions
in section \ref{sec:bh}.

\section{Ricci flat type N spacetimes}

\label{sec:typeN}

\subsection{Overview}

The principal symbol depends on the Riemann tensor of the spacetime. Therefore the simplest non-trivial spacetimes to consider when studying characteristics are those with the simplest non-vanishing Riemann tensor. These are type N spacetimes. In this section we will determine the characteristic hypersurfaces of an arbitrary Ricci flat type N spacetime.

Introduce a null basis $e_0^a \equiv \ell^a$, $e_1^a \equiv n^a$, $e_i^a \equiv m_i^a$ with $\ell$ and $n$ null with $\ell \cdot n = 1$ and $m_i$ a set of orthonormal spacelike vectors orthogonal to $\ell$ and $n$. A spacetime is Ricci flat with a Weyl tensor of type N (in the classification of Ref.~\cite{cmpp}) if, and only if, one can choose the basis so that the only non-vanishing components of the Riemann tensor are (in the notation of \cite{GHP})
\be
\label{typeN}
 \Omega'_{ij} \equiv R_{1i1j} 
\;.\ee
Note that $\Omega'_{ij}$ is a symmetric traceless $(d-2) \times (d-2)$ matrix. It is easy to see that any such spacetime is a solution of Lovelock theory with $\Lambda=0$ (generalising a result of Ref.~\cite{pravda}). 

We will prove the following result below:

\medskip

{\noindent \bf Proposition.} For a generic Ricci flat type N spacetime, there exist $d(d-3)/2$ ``effective metrics"
\be
\label{efftypeN}
 G^{ab}_I = g^{ab} + \omega_I \ell^a \ell^b
\;.\ee
$I = 1, \ldots, d(d-3)/2$ such that a hypersurface is characteristic if and only if its normal $\xi_a$ is null with respect to $G_I^{ab}$ for some $I$. The $\omega_I$ are real homogeneous functions of $\Omega'_{ij}$ of weight 1 and independent of $k_p$ for $p>2$.

This result implies that, for a Ricci flat type N spacetime, the normal cone of a Lovelock theory factorizes into a product of quadratic cones, one for each physical polarisation of the metric. 
The tensors $G_I^{ab}$ are non-degenerate with Lorentzian signature. If we view them as inverse metrics then the associated metrics are\footnote{Note that $G_{Iab}$ is the inverse of $G_I^{ab}$, {\it not} the result of lowering indices of $G_I^{ab}$.} 
\be
 G_{Iab} = g_{ab} - \omega_I \ell_a \ell_b
\;.
\ee
Hence characteristic surfaces are null hypersurfaces of these ``effective" metrics. Bicharacteristic curves are the null geodesics of these metrics. For generic $\Omega'_{ij}$, the quantities $\omega_I$ are distinct and hence so are the metrics $G_{I ab}$.

Generically $\omega_I \ne 0$ for all $I$ so the null cones of the effective metrics do not coincide with the null cone of the physical metric. However, the vector $\ell^a$ is null with respect to any of these metrics so the null cones are all tangent along $\ell^a$. The null cones of $G_{I ab}$ form a nested set: the $I$th cone lies inside the $J$th cone if $\omega_I<\omega_J$ hence the outermost cone is the one corresponding to the effective metric with the most positive $\omega_I$. We will show below that there is generically at least one positive $\omega_I$ and so this outermost cone lies outside the light cone (except where they are tangent along $\ell^a$).

It follows from this result that Lovelock theory is hyperbolic in any Ricci flat type N background. To see this choose a hypersurface $\Sigma$ that is spacelike with respect to the outermost null cone and hence with respect to all of them. Pick a codimension 1 surface $S$ within this hypersurface. Then, for each $I$, there are 2 characteristic surfaces passing through $S$, corresponding to ``outgoing" and ``ingoing" hypersurfaces that are null and normal to $S$ with respect to $G_{I ab}$. Hence $\Sigma$ is Lovelock-spacelike and the theory is hyperbolic on $\Sigma$. Since we can do this everywhere in the spacetime, the theory is hyperbolic everywhere in this spacetime. 

These effective metrics determine the causal properties of gravitational propagation in these spacetimes. Consider a linearized gravitational perturbation with initial data on $\Sigma$ that vanishes outside $S$. Then the solution arising from this data will vanish everywhere outside the ``outermost" characteristic surface emanating from $S$. This surface corresponds to the effective metric $G_{Iab}$ with the most positive $\omega_I$. Therefore it is this effective metric, rather than the physical metric, which determines causal properties of gravitational propagation in this background. This outermost characteristic surface is generically spacelike. 

{\it Example}. It is interesting to examine the form of these effective metrics for a very simple Ricci flat type N spacetime. Consider the plane wave spacetime
\be
 ds^2 = a_{ij} x^i x^j du^2 + 2 du dv + \delta_{ij} dx^i dx^j 
\;,\ee
where $a_{ij}$ is constant and traceless. Choosing $m_i = dx^i$ gives $\Omega'_{ij} \propto a_{ij}$. In this case we have $\ell = \partial/\partial v$. It follows that the $I$th effective metric is
\be
G_{I \mu\nu} dx^\mu dx^\nu = (a_{ij}x^i x^j- \omega_I) du^2 + 2 du dv + \delta_{ij} dx^i dx^j
\;.\ee
Since the $\omega_I$ are functions of $\Omega'_{ij}$, they are constant in this spacetime. Hence we can define $v' = v  - \omega_I u/2$ to obtain
\be
G_{I \mu\nu} dx^\mu dx^\nu = a_{ij}x^i x^j  du^2 + 2 du dv' + \delta_{ij} dx^i dx^j
\;,\ee
which shows that the effective metrics are all isometric to the physical metric. However, the isometry is different for each effective metric. Applying this result to the effective metric with the most positive $\omega_I$ we see that causality of Lovelock theory in this spacetime is equivalent to causality defined by the light cone in an isometric spacetime. 

\subsection{Proof of proposition}

We will now prove the above proposition. We will also give explicit expressions for the $\omega_I$ for the case $d=5$.

In the above null basis, the expression for ${\cal R}(x,\xi)$ simplifies: terms with $p>2$ don't contribute so we have (raising and lowering $i,j$ indices freely)
\be
\label{PtypeN}
 ({\cal R} \cdot t)^\mu{}_\nu =
 -16k_2 \delta^{\mu \rho_1\rho_2 1 i}_{\nu \sigma_1 \sigma_2 0j} \xi_{\rho_1} \xi^{\sigma_1} t_{\rho_2}{}^{\sigma_2} \Omega'_{ij} + \frac{16}{d-2} k_2 \delta^\mu_\nu  \delta^{k \rho_1\rho_2 1 i}_{k \sigma_1 \sigma_2 0j} \xi_{\rho_1} \xi^{\sigma_1} t_{\rho_2}{}^{\sigma_2} \Omega'_{ij}
\;.\ee
Previously we've viewed ${\cal R}(x,\xi)$ as a map from $V_{\rm physical}$ to $V_{\rm transverse}$. But now let's just view it as a map taking symmetric tensors to symmetric tensors. We start by proving

\medskip

\noindent {\bf Lemma}. Viewed as a map from symmetric tensors to symmetric tensors, ${\cal R}(x,\xi)$ is diagonalizable with $2d$ eigenvalues that vanish and $d(d-3)/2$ that generically do not vanish.

\medskip

\noindent {\it Proof.} We already know that ${\cal R}$ is gauge invariant so ``pure gauge" modes $t_{ab} = \xi_{(a} X_{b)}$ are eigenvectors with vanishing eigenvalue. There are $d$ such eigenvectors.  Next, if we take $t_{ab} = \ell_{(a} V_{b)}$ for some $V_b$ (i.e.\ the only non-vanishing components are $t_{1\mu}$) then from (\ref{PtypeN}) we have ${\cal R} \cdot t = 0$ so $t$ is an eigenvector with zero eigenvalue. We've already accounted for the case $V_b \propto \xi_b$ (a pure gauge mode), so this gives an additional $d-1$ independent eigenvectors here. So we've identified $2d-1$ of the eigenvectors corresponding to a vanishing eigenvalue, the final one will be determined below. 

Equation (\ref{typeN}) depends only on the choice of $\ell$, the choice of the other basis vectors is arbitrary. In general, $\xi$ will not be parallel to $\ell$ so we can choose $n$ to be a linear combination of $\xi$ and $\ell$. It then follows that $t_{ab} \propto n_{(a} V_{b)}$ is an eigenvector of ${\cal R}$ with eigenvalue $0$ (since it is a linear combination of eigenvectors of the form just discussed). The only non-vanishing components of such $t_{ab}$ are $t_{0\mu}$. 

Now consider an eigenvector $t_{ab}$ of ${\cal R}$ with non-vanishing eigenvalue. Since ${\cal R}$ is symmetric with respect to the inner product (\ref{innerproduct}), $t_{ab}$ must be orthogonal to the eigenvectors with vanishing eigenvalue just discussed. This implies that its only non-vanishing components are $t_{01}$ and $t_{ij}$ with $t_{ii}=0$. Hence $t_{ab}$ has the form
\be
\label{nonzeroevec}
 t_{ab} = 2  t_{01} \ell_{(a} n_{b)} + t_{ij} m_{ia} m_{ib} \qquad t_{ii}=0
\;,\ee
which obeys the transversality condition (\ref{transverse}). The first term belongs to the kernel of ${\cal R}$ so ${\cal R} \cdot t$ is independent of $t_{01}$. Consider the eigenvalue equation ${\cal R} \cdot t = \lambda t$. Since $\lambda \ne 0$, the $01$ component fixes $t_{01}$ in terms of $t_{ij}$. The $0i$ and $1i$ components are trivial (because ${\cal R} \cdot t$ is orthogonal to the eigenvectors of ${\cal R}$ with vanishing eigenvalue) and so only the ${ij}$ components remain. These give 
\be
 ({\cal R} \cdot t)_{ij} =  
16k_2
\xi_0^2{\cal O}(t)_{ij}
 \;,\ee
where, for a traceless symmetric matrix $t_{ij}$ we define
\be
 {\cal O}(t)_{ij}=
t_{ik} \Omega'_{kj} + t_{jk} \Omega'_{ki} - \frac{2}{d-2}t_{kl} \Omega'_{kl} \delta_{ij} 
\;.\ee
Hence to determine the non-vanishing eigenvalues of ${\cal R}$ we need to find the eigenvalues of ${\cal O}$. Let ${\cal V}$ be the space of $(d-2) \times (d-2)$ traceless symmetric matrices, which has dimension $d(d-3)/2$. ${\cal O}$ maps ${\cal V}$ to itself. If $X,Y$ are both traceless symmetric then $Y_{ij} {\cal O}(X)_{ij} = X_{ij} {\cal O}(Y)_{ij}$ so ${\cal O}$ is symmetric with respect to the Euclidean metric on ${\cal V}$. Hence ${\cal O}$ has real eigenvalues $\nu_I$ and the associated eigenvectors form a basis for ${\cal V}$. The eigenvalues $\nu_I$ are homogeneous functions of $\Omega'_{ij}$ of weight 1. 

We have shown that ${\cal R}$ has eigenvalues $-(1/2) \xi_0^2 \omega_I$ where
\be
 \omega_I = -
32k_2
\nu_I\;, \qquad I = 1,2,  \ldots , d(d-3)/2\;,
\ee
and the corresponding eigenvectors are linearly independent and of the form (\ref{nonzeroevec}). 

The final eigenvector of ${\cal R}$ must have vanishing eigenvalue. This can be found by taking $t_{ab}$ to have $t_{0\mu} = t_{1\mu}=0$ and decomposing $t_{ij} = \hat{t}_{ij} + \alpha \delta_{ij}$ where $\hat{t}_{ij}$ is traceless and can therefore be expanded in terms of the basis of eigenvectors of ${\cal O}$ just discussed. If the eigenvalues of ${\cal O}$ are all non-zero then ${\cal R} \cdot t = 0$ can then be solved to determine $\hat{t}_{ij}$ uniquely in terms of $\alpha$. $\Box$

We can determine the eigenvalues $\nu_I$ more explicitly by choosing the spatial basis vectors $m_i$ to diagonalize $\Omega'_{ij}$. The diagonal elements are the eigenvalues of $\Omega'_{ij}$, which we denote by $\Omega_{(i)}'$. Consider a traceless symmetric matrix $t_{kl}$ for which only the $\{i,j\}$ components are non-vanishing with $i \ne j$. This is an eigenvector of ${\cal O}$ with eigenvalue $\nu_I = \Omega'_{(i)} + \Omega'_{(j)}$. There are $(1/2) (d-2)(d-3)$ such eigenvectors. Since the sum of the $\Omega'_{(i)}$ is zero, so must be the sum of these $\nu_I$. Hence generically some of them are positive and some are negative.

To determine the remaining $d-3$ eigenvalues, consider a traceless symmetric matrix $t_{kl}$ for which only the diagonal components are non-vanishing. For such a matrix, the eigenvalue equation for ${\cal O}$ reduces to finding the eigenvalues of a $(d-3) \times (d-3)$ symmetric matrix. For $d=5$ we can do this explicitly with the result
\be
\label{special5d}
\nu_I = \pm \sqrt{\frac{2}{3} \Omega'_{ij} \Omega'_{ij}}\;, \qquad d=5
\;.\ee
We discussed above how the effective metric $G_{Iab}$ corresponding to the most positive $\omega_I$ determines the causal properties of Lovelock theory in this spacetime. For $d=5$, it is easy to show that the most positive $\omega_I$ corresponds to one of the roots in (\ref{special5d}), which is always non-zero (unless the spacetime is flat). 

Armed with this Lemma we can now prove the above proposition. 

\noindent {\it Proof of proposition.} Let $\xi_a$ be normal to a characteristic hypersurface. If $\xi_a$ is non-null then from section \ref{sec:lovelockcharacteristic} we know that the characteristic condition reduces to the existence of non-zero $\hat{t}_{ab} \in V_{\rm transverse}$ obeying (\ref{evaleq}), i.e., $\hat{t}_{ab}$ is an eigenvector of ${\cal R}(x,\xi)$ with non-zero eigenvalue. We determined these above: they are the $d(d-3)/2$ eigenvectors of the form (\ref{nonzeroevec}) with eigenvalue $-(1/2) \xi_0^2 \omega_I$. Hence from (\ref{evaleq}), a non-null $\xi_a$ is characteristic if and only if, for some $I$,
\be
 \xi^2 = -\xi_0^2 \omega_I = - (\ell^a \xi_a)^2 \omega_I 
\ee
which can be rewritten as (\ref{efftypeN}). So for a generic type N spacetime we have $d(d-3)/2$ non-null characteristic $\xi_a$. 

Now assume that $\xi_a$ is null and not parallel to $\ell_a$. In this case we can choose our second basis vector $n_a = \xi_a$ (rescaling $\xi_a$ as required). The conditions for null $\xi_a$ to be characteristic were given in section \ref{sec:lovelockcharacteristic}. In that section we chose $e_0 = \xi$ but above we chose $e_0 = \ell$ and $e_1 = n$. We will adopt the convention of section \ref{sec:lovelockcharacteristic} so we will need to swap $0$ and $1$ indices when using results from the above Lemma. Using the Lemma, we can expand $t_{ab}$ in terms of the eigenvectors of ${\cal R}$. Equations (\ref{ijcpt}) and (\ref{nonzeroevec}) then imply that the coefficients in this expansion of the eigenvectors corresponding to non-zero eigenvalues must vanish. So $t_{ab}$ must belong to the kernel of ${\cal R}$, i.e., all the ${\cal R} \cdot t$ terms vanish in equations \eqref{cpt} which implies $t_{00} = t_{0i} = t_{ii} = 0$. Components of the form $t_{1\mu}$ are pure gauge. This leaves only components of the form $t_{ij}$ where $t_{ii}=0$. But these are precisely the eigenvectors (\ref{nonzeroevec}) which generically correspond to non-zero eigenvalue, which we've already excluded. So only when any of the eigenvalues $\omega_I$ happens to vanish is it possible for a characteristic direction $\xi_a$ to be null and not parallel to $\ell_a$.

Finally, if $\xi_a$ is null and parallel to $\ell_a$ then ${\cal R}(x,\xi)$ vanishes so the analysis is the same as for GR (section \ref{sec:GR}). Such $\xi_a$ is characteristic and non-trivial elements of the kernel of $P$ correspond to the subspace $V_{\rm physicaltransverse}$ of $V_{\rm physical}$, with dimension $d(d-3)/2$. This is in agreement with the proposition since $\ell_a$ is null with respect to all of the $G_I^{ab}$. 

\section{Static, maximally symmetric black holes}
\label{sec:bh}
\subsection{General properties and effective metrics}
In this section we will determine the characteristic hypersurfaces of certain black hole solutions.  Lovelock theories admit static, maximally symmetric black hole solutions with metric of the form \cite{boulware,wheeler,cai,cai2}
\be
 ds^2 = -f( r) dt^2 + f( r)^{-1} dr^2 + r^2 d\Sigma^2
\ee
or, using a tortoise coordinate $r_*$ such that $dr_* = dr/f(r)$,
\be
 ds^2 = f( r) \left( - dt^2 + dr_*^2 \right) + r^2 d\Sigma^2\;,
\ee
where $d\Sigma^2$ is the line element of a $d-2$ dimensional space ${\cal S}$ with constant curvature of sign $\kappa =0$, $1$, or $-1$. Defining the function
\be
f(r)=\kappa-r^2\psi(r)\;,
\ee
these solutions satisfy the algebraic relation
\be\label{algrel}
W[\psi]\equiv-\sum_{p\geq2}\left[2^{p+1}k_p\left(\prod_{k=1}^{2p-2}(d-2-k)\right)\psi^p\right]+\psi-\frac{2\Lambda}{(d-1)(d-2)}=\frac{\mu}{r^{d-1}}\;,
\ee
where the constant $\mu$ is proportional to the ADM mass.  

In this section, we will determine the characteristic hypersurfaces of Lovelock theories in these spacetimes. To do this we need the form of the Riemann tensor, which is given e.g.\ in \cite{deser}. In an orthonormal basis $e_0=  -f^{1/2} dt$, $e_1 = f^{-1/2} dr$, $e_i$ tangent to $d\Sigma$, the non-vanishing Riemann components are of the form
\bea
  R_{IJKL} &=& R_1 ( r) \left( \eta_{IK} \eta_{JL} - 
\eta_{IL} 
\eta_{JK} \right)\;, \qquad R_{IiJj} = R_2 ( r)\eta_{IJ} \delta_{ij}\;, \nonumber \\ R_{ijkl} &=& R_3( r) \left( \delta_{ik} \delta_{jl} - \delta_{il} \delta_{jk} \right)
\;,\eea
where indices $I,J$ take values in $\{0,1\}$ and $i,j$ take values in $\{2, \ldots , d-1 \}$ and $\eta_{IJ}$ is the 
two-dimensional
Minkowski metric. Note that the Riemann tensor at $p$ is invariant under a subgroup of the Lorentz group acting on the tangent space at $p$, consisting of 
two-dimensional
Lorentz boosts acting on the $IJ$ indices and rotations acting on the $ij$ indices. 

The characteristic determinant $Q$ is a polynomial in $\xi_\mu$:
\be
 Q = Q^{\mu_1 \ldots \mu_N} (r) \xi_{\mu_1} \ldots \xi_{\mu_N}
\;,\ee
where $N = d(d-3)$. The coefficients $Q^{\mu_1 \ldots \mu_N}$ will inherit the symmetry of the metric and Riemann tensor, which implies that they are functions of $r$ times products of $\eta^{IJ}$ and $\delta^{ij}$. Hence $Q$ depends on $\xi_\mu$ only in the combinations $\eta^{IJ} \xi_I \xi_J$ and $\delta^{ij} \xi_i \xi_j$. 

We've shown that $Q=Q(r,\xi^I \xi_I,\xi^i \xi_i)$. Since this is a homogeneous polynomial of degree $N$ in $\xi_\mu$, we can divide through by $(\xi^i \xi_i)^{N/2}$ to obtain a polynomial in $\xi^I \xi_I /\xi^i \xi_i$ which must have one or more real roots $-c_A( r)$ if non-trivial characteristic surfaces exist. Hence we can factorize $Q$ into a product of factors of the form $(\xi^I \xi_I + c_A ( r) \xi^i \xi_i)^{p_A}$ and (possibly) a factor without any real roots.\footnote{We will argue below that a factor without real roots does not occur.} We can write the former as $(G_A^{ab} \xi_a \xi_b)^{p_A}$ where
\be
\label{effectivespherical}
 G_A^{IJ} = \eta^{IJ} \qquad G_A^{Ii} = 0 \qquad G_A^{ij} = c_A(r) \delta^{ij} 
\;.\ee
Therefore, just as the type N case studied above, a hypersurface is characteristic if, and only if, it is null with respect to one of the ``effective metrics" $G_A^{ab}$. Note that $G_A^{ab}$ is Lorentzian if $c_A(r)>0$, degenerate if $c_A(r)=0$, and Lorentzian with ``mostly minus" signature if $c_A(r)<0$. If $c_A(r) \ne 0$ then we can define $G_{Aab}$ as the inverse of $G_A^{ab}$ ({\it not} by lowering indices on $G_A^{ab}$) which gives
\be
 G_{A\mu\nu} dx^\mu dx^\nu = -f(r) dt^2 + f(r)^{-1} dr^2 + \frac{r^2}{c_A(r)} d\Sigma^2
\;.\ee

Rather than compute the effective metrics directly from the Riemann tensor, we can make use of results on linear perturbations of these spacetimes \cite{tensorperts,tensorperts2,scalarperts,Konoplya:2008ix,takahashi,takahashi2,Takahashi:2010ye,Takahashi:2010gz}. If we linearize around a solution then the term involving second derivatives of $\delta g_{\mu\nu}$ is
\be
 \frac{\delta E_{\mu\nu}}{\delta (\partial_\alpha \partial_\beta g_{\rho \sigma})} \partial_\alpha \partial_\beta \delta g_{\rho \sigma}
\;,\ee
where $E_{\mu\nu}=0$ is the equation of motion. The coefficient here is the same matrix which, when contracted with $\xi_\alpha \xi_\beta$ gives the principal symbol. Hence we can determine the principal symbol by looking at the second-derivative terms in the equations of motion for linearized perturbations. 

Linearized perturbations are studied by decomposing perturbations into scalar, vector and tensor types with respect to $\mathcal S$, and then expanding these in harmonics on ${\cal S}$. This leads to a single ``master equation" for each type of perturbation, which can be written as a 2d wave equation with a potential:
\be
\left(  -\frac{\partial^2}{\partial t^2} + \frac{\partial^2}{\partial r_*^2} -V_l (r) \right) \Psi_l(t,r) = 0
\;,\ee
where $r_*$ is a tortoise coordinate ($dr_* = dr/f(r)$) and the parameter $l$ labels the harmonic (e.g.\ $l=2,3,\ldots$ for spherical symmetry). To determine the principal symbol, we need to ``undo" the expansion in harmonics so that we can read off the terms involving second derivatives on $\mathcal S$. We can do this by considering perturbations that oscillate very rapidly so that the second derivatives dominate the equation. Rapid oscillation corresponds to large $l$. At large $l$, the harmonics satisfy $D^2 Y^{(l)} \approx -l^2 Y^{(l)}$ where $D^2$ is the Laplacian on ${\cal S}$. At large $l$, the potential obeys $V_l(r) \approx l^2 f(r)c_A(r)/r^2$ for some function $c_A(r)$ where the index $A \in \{ {\rm S,V,T} \}$ (scalar, vector, tensor). We deduce that the term involving second derivatives on ${\cal S}$ must be $f(r)c_A(r) D^2/r^2$. Therefore the second derivative terms in the equation for linearized perturbations of a given type (scalar, vector or tensor) are
\be
 \left(  -\frac{\partial^2}{\partial t^2} + \frac{\partial^2}{\partial r_*^2}+\frac{f(r) c_A(r) D^2}{r^2} \right) \Psi \equiv f(r) G_A^{\mu\nu} \partial_\mu \partial_\nu \Psi
\ee
from which we can read off the components of the effective metric $G_A^{\mu\nu}$, and see that it takes the form (\ref{effectivespherical}). Note that we've made a particular choice for the overall conformal factor in the effective metric (the factor of $f(r)$ above). This is purely for convenience: it does not affect the definition of characteristic hypersurfaces since these are null with respect to the effective metric.

We've shown that for each type of perturbation (scalar, vector, tensor) we can define an effective metric and the function $c_A(r)$ is determined by the large $l$ behaviour of the potential in the master equation for linearized perturbations of that type. We will give results for $c_A( r)$ below but first we will make some more general observations.  

Since scalar, vector and tensor perturbations form a basis for all perturbations, it follows that they exhaust the physical degrees of freedom of the graviton, so the characteristic determinant must factorize fully:
\be
 Q(x,\xi) = (G_{S}^{ab}(x) \xi_a \xi_b)^{p_S} (G_{V}^{cd}(x) \xi_c \xi_d)^{p_V}
  (G_{T}^{ef}(x) \xi_e \xi_f)^{p_T} 
\ee 
where $p_S$, $p_V$, $p_T$ are the number of degrees of freedom of scalar, vector and tensor perturbations respectively, so $p_S+p_V+p_T = d(d-3)/2$.

For a black hole that is large (compared to the length scales set by the constants $k_p$), the functions $c_A( r)$ are positive everywhere outside the event horizon (we will not discuss the black hole interior). Let's discuss this case first. The effective metrics $G_{Aab}$ are smooth and Lorentzian. The characteristic surfaces are the null hypersurfaces of these effective metrics, so the null cones of $G_{Aab}$ determine causality of Lovelock theory in this spacetime. 

If a vector $V^a$ is timelike with respect to $G_{Aab}$ then it is also timelike with respect to $G_{A'ab}$ if $c_{A'}(r )>c_A(r)$. Hence the null cones at a point $p$ form a nested set with the innermost cone corresponding to the smallest $c_A(r)$ and the outermost cone to the largest $c_A( r)$. The null cone of the physical metric can be included in this nested set. If $c_A (r )>1$ then the outermost graviton cone lies outside the light cone: gravity travels faster than light. We will see below that this is often the case, including arbitrarily far from the black hole. 

\begin{figure}[htbp]
\subfigure[$r=1.5$]{\includegraphics[width=5cm,clip]{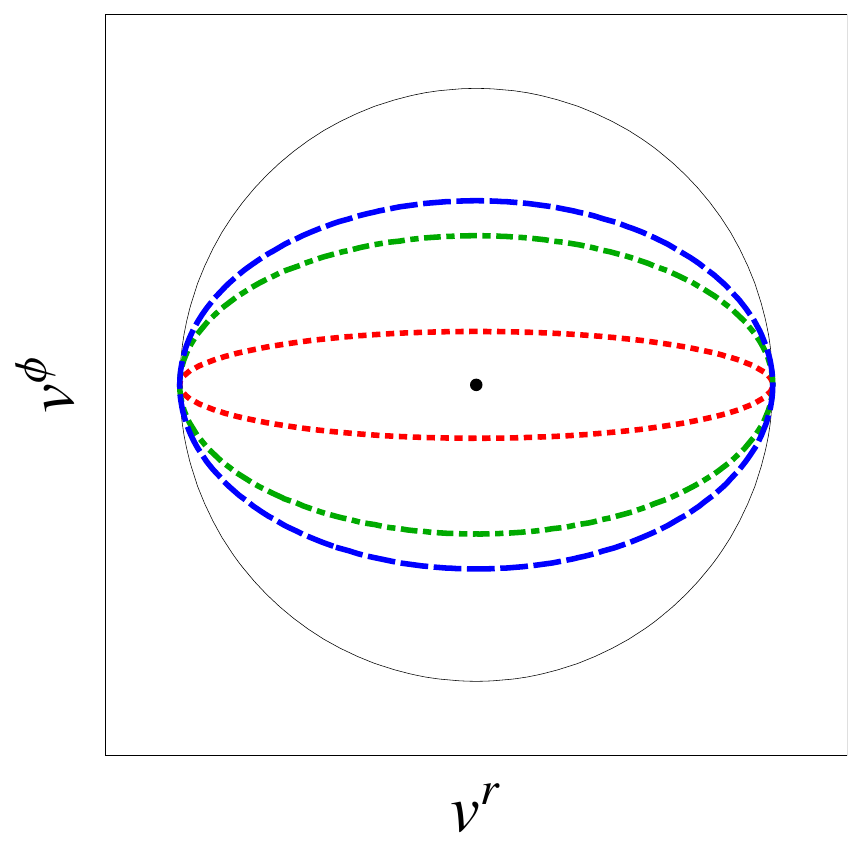}}
\subfigure[$r=3.0$]{\includegraphics[width=5cm,clip]{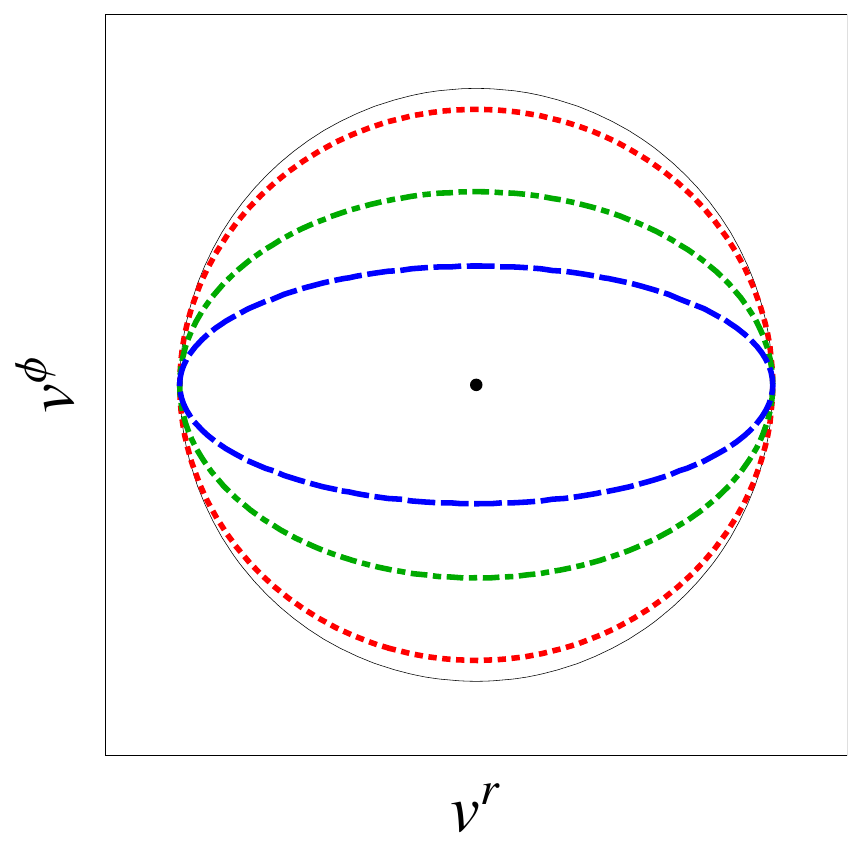}}
\subfigure[$r=4.0$]{\includegraphics[width=5cm,clip]{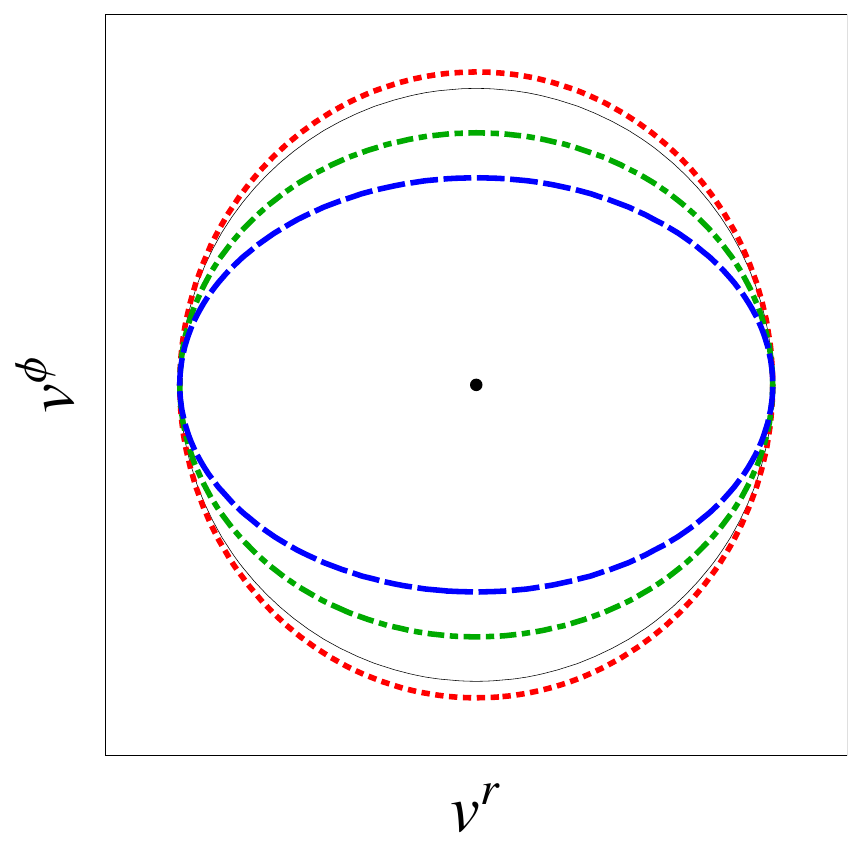}}
\caption{Cross-section of null cones of the effective metrics, and the light cone of the physical metric, with axes scaled so that the light cone appears as a circle.  The solid black curve shows the light cone with respect to the physical metric. The dotted red curve, the dot - dashed green curve, and the dashed blue curve give the null cones for tensor, vector, and scalar perturbation sectors, respectively. Here we give results for a spherically symmetric solution ($\kappa=1$) with $\Lambda=0$, $r_0=1$, $k_2=-1/4$ in $d=7$ at $r=1.5, 3, 4$. We consider a vector lying in the equatorial plane, with components $v^t,v^r,v^\phi$ (with $\phi \sim \phi+2\pi$ an angular coordinate). The plots show a cross-section of the null cones with $v^t = f^{-1}$, which implies $(v^r)^2 +f r^{-2} c_A^{-1}(v^\phi)^2=1$.
}
\label{Fig:cones}
\end{figure}

In general, these null cones (including that of the physical metric) coincide for vectors orthogonal to ${\cal S}$, i.e.\ for radial null vectors. Hence we recover the result that ``gravity travels at the speed of light on the radial direction" \cite{brigante2}. So the null cones are all tangent along two lines corresponding to the ingoing and outgoing radial directions, but otherwise they are distinct. Fig.~\ref{Fig:cones} shows cross-sections through the different null cones 
for different values of $r$ in a certain spherically symmetric black hole spacetime. We will study properties of these cones in more detail in the following sections.

Causality is determined by the outermost null cone. This need not be the same everywhere in the spacetime. For the  example in Fig. \ref{Fig:cones}, the outermost null cone corresponds to the tensors at large $r$ but to the scalars at smaller $r$. 

For these solutions with all $c_A$ positive, it is clear that we can find Lovelock-spacelike hypersurfaces, which are hypersurfaces that are spacelike with respect to all of the $G_{Aab}$. For example, surfaces of constant $t$ have this property. Hence  Lovelock theory is hyperbolic in these backgrounds.

The situation changes dramatically for some small Lovelock black holes. In this case, the functions $c_A ( r)$ are all positive far from the black hole but for some small black holes there is a critical radius $r_c>r_0$ (where $r=r_0$ is the event horizon) such that one of the $c_A$ vanishes at $r=r_c$ and becomes negative for $r<r_c$. The other $c_A( r)$ remain positive. (For spherical black holes in Einstein-Gauss-Bonnet theory with $\Lambda=0$ this happens for scalars when $d=5$ and for tensors when $d=6$.) Call the modes (scalar, vector or tensor) for which $c_A$ changes sign the ``bad" modes and the others the ``good" modes. 

This phenomenon implies that Lovelock theory is not hyperbolic for $r \le r_c$ in these spacetimes. For example, surfaces of constant $t$ are spacelike for the good modes but timelike in $r<r_c$ for the bad modes. This implies that the problem of studying stability by specifying an initial perturbation on a surface of constant $t$ and evolving in time is ill-posed. It seems likely that there will be serious problems at the nonlinear level when the "good" and "bad" modes are coupled together. But the problem is ill-posed even at the linearized level, as we will now explain. 
 We will show that solutions of the linearized equation of motion for the "bad" modes do not depend continuously on the initial data for such perturbations on a surface $t=0$. We will also argue that, for a generic smooth initial perturbation, a solution of the linearized equation of motion does not even exist.

Consider a solution of the master equation of the form $e^{-i\omega t} \chi_l(r_*)$. The master equation reduces to the Schrodinger equation with potential $V_l(r_*)$ and energy $\omega^2$. But for the bad modes, $V_l( r)$ is negative for $r \le r_{\rm bad}$, and admits negative energy bound states \cite{tensorperts}.
 Since $V_l$ scales as $l^2$, the energy $\omega^2$ of the bound state must scale as $-\alpha^2 l^2$ for large $l$ where $\alpha$ is a constant. Hence for large $l$ there exist solutions of the master equation of the form $\Psi_l(t,r) = e^{\alpha l t} \chi_l(r_*)$ where $\chi_l(r_*)$ is the bound state wave function (which we assume to be normalised). These grow exponentially in time and are regular on the future horizon.\footnote{To see this, note that $V_l(r_*)$ vanishes for $r_* \rightarrow \pm \infty$ hence $\chi_l(r_*) \propto e^{\pm i \omega r_*} \sim e^{\mp \alpha l r_*}$ as $r_* \rightarrow \pm \infty$. Bound states must decay for $r_* \rightarrow \pm \infty$ hence $\chi_l(r_*) \sim e^{\alpha l r_*}$ as $r_* \rightarrow -\infty$ so $e^{\alpha l t} \chi_l(r_*) \propto e^{\alpha l v}$ as $r_* \rightarrow -\infty$ where $v= t+r_*$ is the ingoing Eddington-Finkelstein coordinate, which is regular on the future horizon.} Such solutions have been interpreted previously as an instability of the black hole.
 
For simplicity, consider the case for which the "bad" modes are the tensors (the other cases are completely analogous). Then the solution $\Psi_l(t,r)$ corresponds to a linearized metric perturbation of the form $r^p \Psi_l(t,r) Y^{(l)}_{\mu\nu}(x)$ for some $p$, where $x$ denotes the coordinates on ${\cal S}$ and $Y^{(l)}_{\mu\nu}$ is a tensor harmonic on ${\cal S}$. Now let $\delta g^{(l)}_{\mu\nu}(t,r,x) = e^{-\sqrt{l}} r^p \Psi_l(t,r) Y^{(l)}_{\mu\nu}(x)$. This defines a sequence of solutions of the linearized equation of motion. The initial data at $t=0$ is $\delta g^{(l)}_{\mu\nu}(0,r,x)$ and $\partial_t \delta g^{(l)}_{\mu\nu}(0,r,x)$. Now take the limit $l \rightarrow \infty$. The factor $e^{-\sqrt{l}}$ ensures that the initial data, and all of its derivatives (w.r.t. $r,x$) vanishes in this limit. Therefore, if the solution depends continuously on the initial data, it should vanish as $l \rightarrow \infty$ for $t>0$. But this is not the case: for large $l$ we have $\delta g^{(l)}_{\mu\nu} \propto e^{-\sqrt{l}} e^{\alpha l t}$ which diverges as $l \rightarrow \infty$ for any $t>0$. Hence solutions of the linearized equation for the bad modes do not depend continuously on the initial data. 

We can also argue that for generic initial data there exists no solution at all for arbitrarily small $t>0$. Any smooth initial data can be decomposed into harmonics at $t=0$ and expanded in a basis of modes of the form $\delta g_{\mu\nu}^{(l)}$. At $t=0$ there will be no problem with the decomposition into harmonics. But for any $t>0$ the factor $e^{\alpha l t}$ implies that the sum over $l$ generically will not converge. Only for very special initial data (e.g. analytic initial data) will a solution of the linearized equations exist locally.

\subsection{Effective metrics in Einstein-Gauss-Bonnet}
We will now describe the functions $c_A( r)$ in more detail.   Using results from the master equations \cite{Takahashi:2010ye}, these are given 
for general Lovelock theory
by
\begin{subequations}\label{ceq}
\begin{align}\label{cteq}
c_T(r)&=-\left(1+\frac{1}{d-4}\right)A(r)-\left(1-\frac{1}{d-4}\right)\frac{1}{A(r)}+B(r)+3\\\label{cveq}
c_V(r)&=A(r)\\\label{cseq}
c_S(r)&=3\left(1-\frac{1}{d-2}\right)A(r)+\left(1-\frac{3}{d-2}\right)\frac{1}{A(r)}-\left(1-\frac{2}{d-2}\right)(B(r)+3)\;,
\end{align}
\end{subequations}
where 
\begin{subequations}\label{ABeq}
\begin{align}\label{Aeq}
A(r)&=1-\frac{d-1}{d-3}\frac{W''[\psi(r)]W[\psi(r)]}{W'[\psi(r)]^2}\\\label{Beq}
B(r)&=\frac{(d-1)^2}{(d-3)(d-4)}\frac{W[\psi(r)]^2W'''[\psi(r)]}{A(r)W'[\psi(r)]^3}\;.
\end{align}
\end{subequations}

In the rest of this subsection, we will focus on Einstein-Gauss-Bonnet theory, defined by $k_p=0$ for $p>2$.  We will allow for a cosmological constant. The algebraic relation \eqref{algrel} can be solved and gives \cite{boulware}
\be
f(r)=\kappa+\frac{r^2(1-q(r))}{\tilde\alpha_2}\;,\qquad q(r)=\sqrt{1+2\tilde\alpha_2\left(\frac{\mu}{r^{d-1}}+\frac{2\Lambda}{(d-1)(d-2)}\right)}
\;,\ee
where $\tilde\alpha_2=-16k_2(d-3)(d-4)$, and we have chosen the branch of solutions that are asymptotically flat when $\Lambda=0$.  
In order for this to make sense for $\mu=0$, we must have 
\be
-\frac{1}{4}(d-1)(d-2)<\tilde\alpha_2\Lambda\;.
\label{bound}
\ee
In this case 
of Einstein-Gauss-Bonnet theory,
we have
\be
A(r)=\frac{1}{q(r)^2}\left[\frac{1}{2}+\frac{1}{d-3}\left(1+\frac{2\tilde\alpha_2\Lambda}{d-2}\right)\right]+\left(\frac{1}{2}-\frac{1}{d-3}\right)\;,\qquad B(r)=0\;.
\ee
Let $r=r_0$ denote the radius of the event horizon
given by $f(r_0)=0$. 
We will use $r_0$ rather than $\mu$ as a parameter. 

Instead of plotting $c_A( r)$ we will show plots of
\be
 V_{eff}(r ) = \frac{f( r) c_A( r)}{2r^2}
\ee
which is the effective potential for null geodesics of the metric $G_{A\mu\nu}$ (which are bicharacteristic curves), along with the corresponding effective potential for null geodesics of the physical metric (obtained by setting $c_A = 1$). 

\begin{figure}[th]
\begin{center}
\includegraphics[width=.45\textwidth]{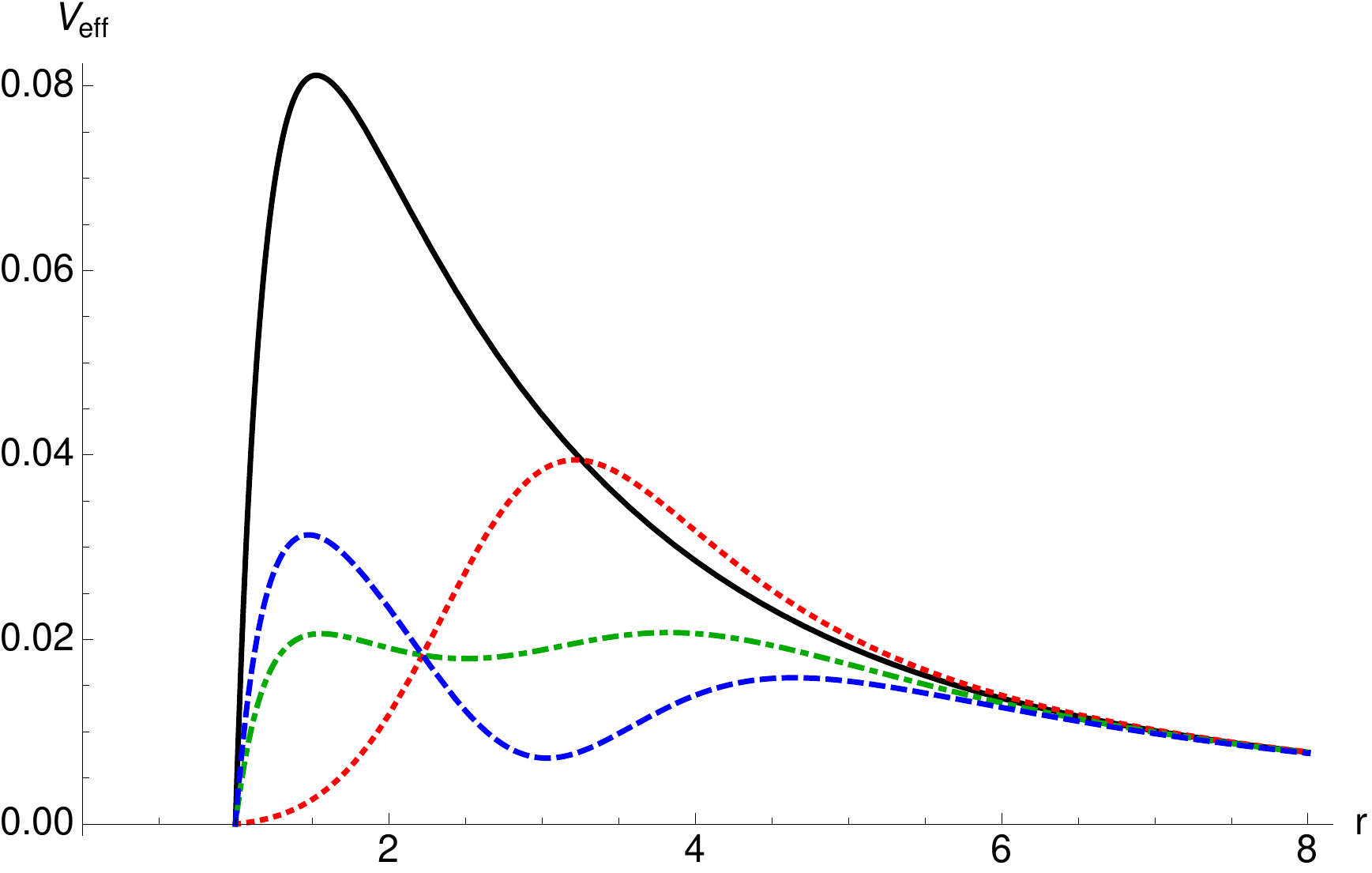}
\end{center}
\caption{Effective potentials for $\Lambda=0$, $\kappa=1$, $r_0=1$, $k_2=-1/4$ in $d=7$. The solid black curve corresponds to the physical metric.  The dotted red curve, the dot - dashed green curve, and the dashed blue curve give the tensor, vector, and scalar perturbation sectors, respectively.}\label{Fig:Veff1}
\end{figure}  
\begin{figure}[th]
\begin{center}
\includegraphics[width=.45\textwidth]{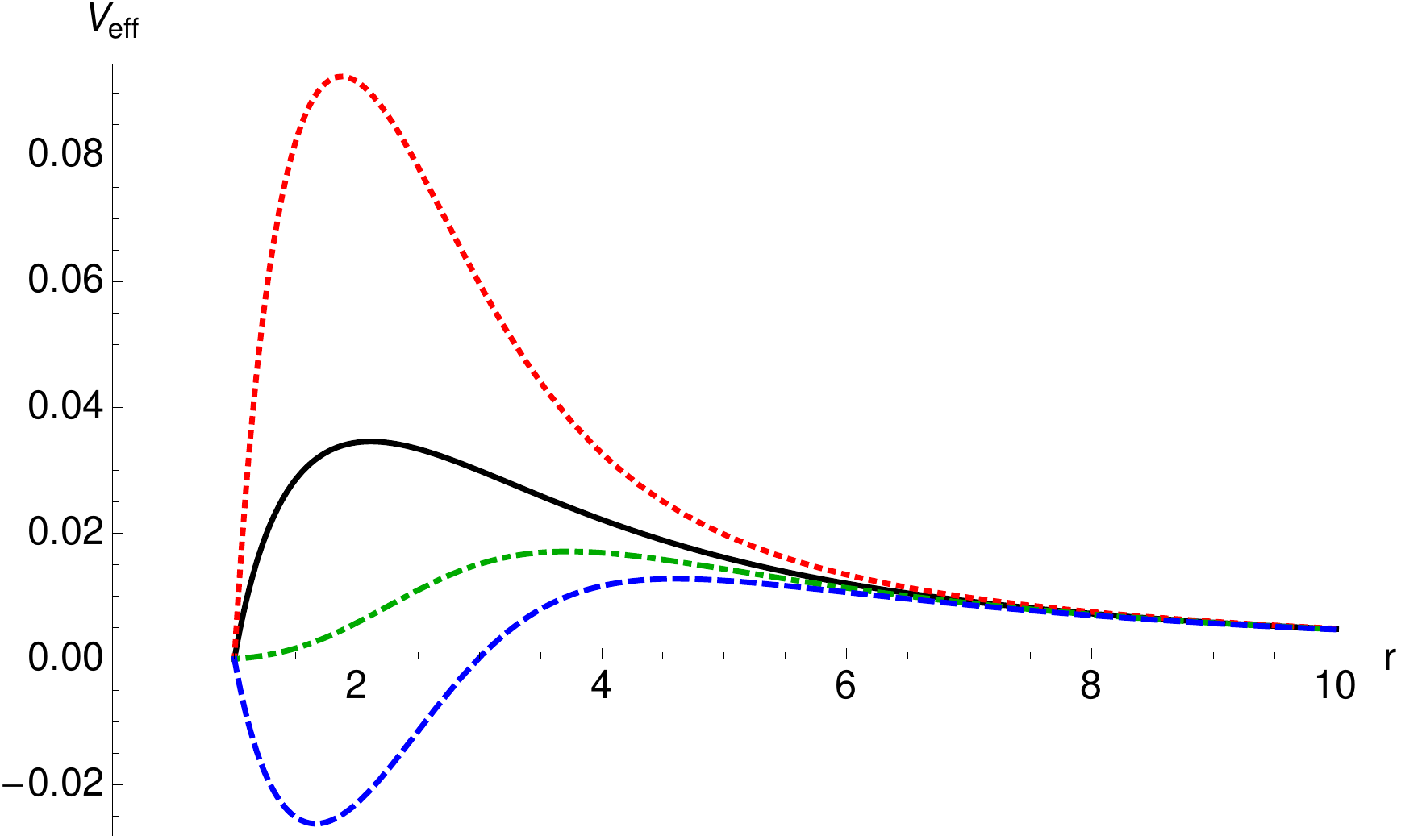}\qquad
\includegraphics[width=.45\textwidth]{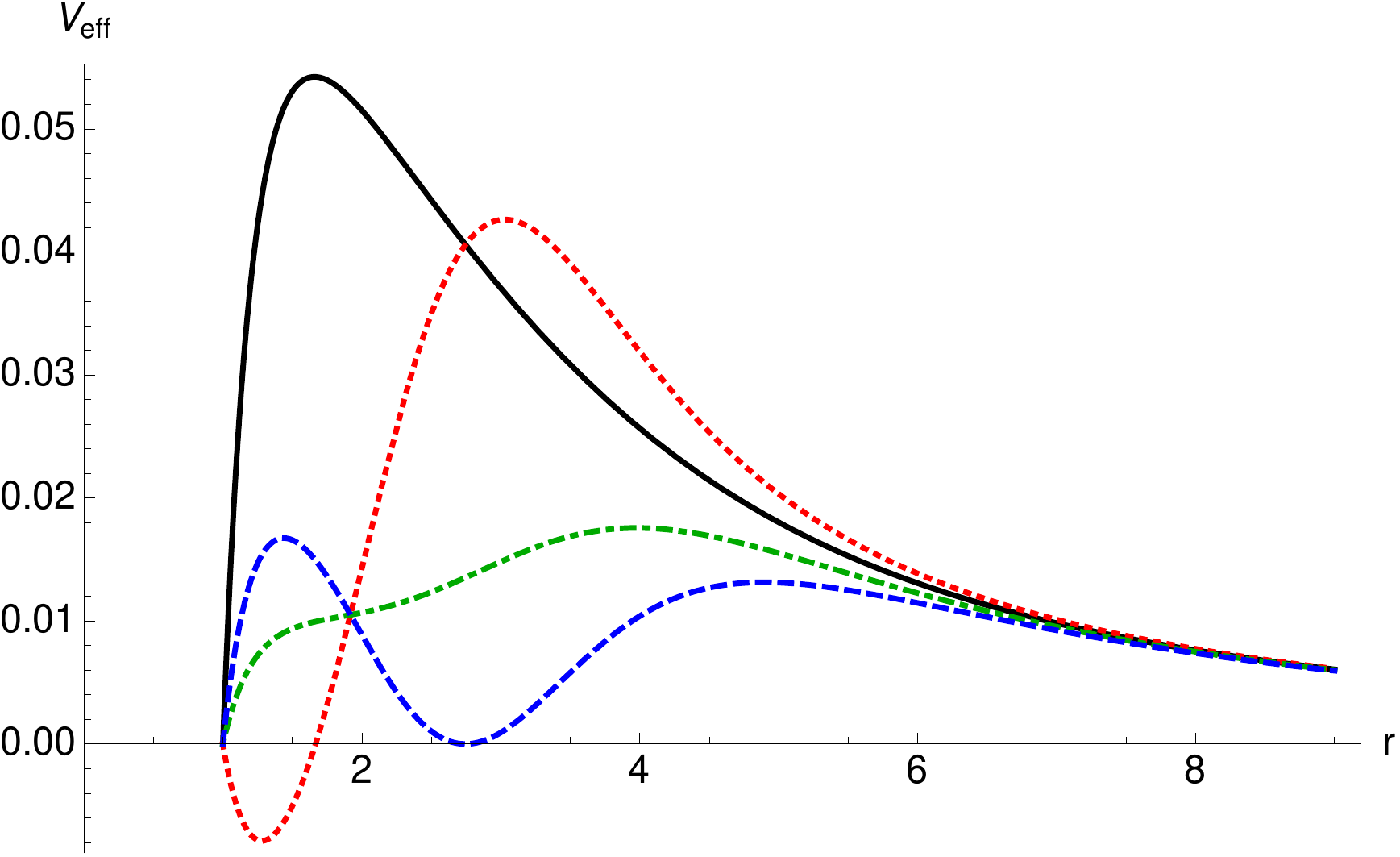}
\end{center}
\caption{Effective potentials for $\Lambda=0$, $\kappa=1$, $r_0=1$, $k_2=-1/4$ in $d=5$ (left) and $d=6$ (right).  Same colour scheme as figure \ref{Fig:Veff1}.}\label{Fig:Veff2}
\end{figure}  
\begin{figure}[th]
\begin{center}
\includegraphics[width=.45\textwidth]{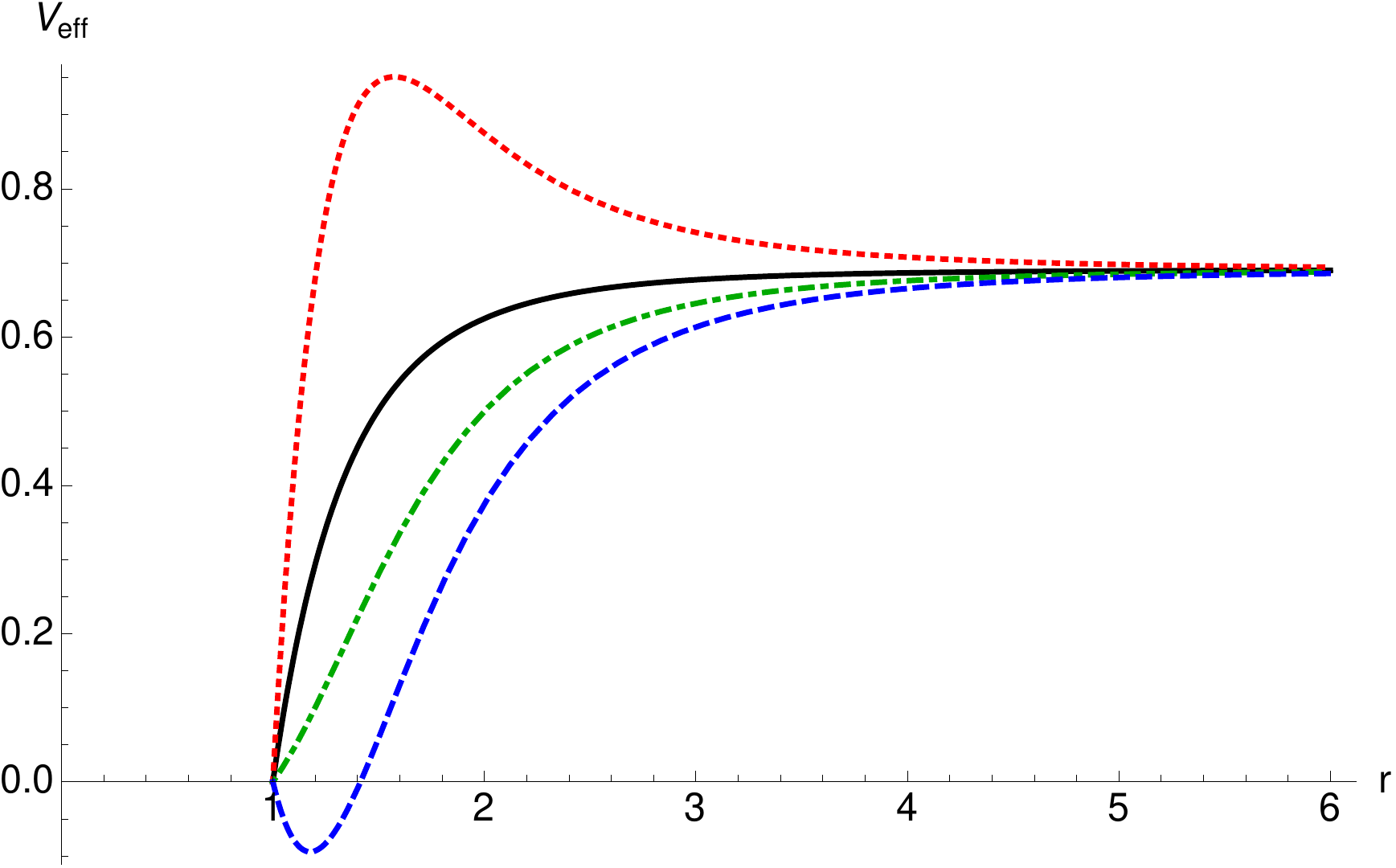}
\end{center}
\caption{Right: effective potential for AdS length $\ell=1$, $\kappa=0$, $r_0=1$, $k_2=-1/80$ in $d=5$.  Same colour scheme as figure \ref{Fig:Veff1}.}\label{Fig:Veffads}
\end{figure}  
Figure \ref{Fig:Veff1} shows the behaviour of the effective potentials for a small spherical ($\kappa=1$) black hole with $d=7$ and $\Lambda=0$. From the plot we see that $0<c_S<c_V<1<c_T$ for large $r$. Hence at large $r$, the effective metric that determines causality is that of the tensors, and these modes propagate faster than light. At intermediate $r$ we have $0<c_S<c_V<c_T<1$ so it is still the tensor metric that determines causality but now all modes propagate slower than light.\footnote{
As $r$ approaches the horizon, $c_T=c_V=c_S<1$ is realized at a point where
$A(r) = \frac{d-5}{2d-7}$.
}
Near to the horizon, we have $0<c_T<c_V<c_S<1$ and so it is now the effective metric for the scalars that determines causality, with all modes subluminal. Since $c_A>0$ for all $A$, the theory is hyperbolic in this spacetime. The results of Ref.~\cite{Konoplya:2008ix} suggest that this behaviour extends to $d \ge 7$. 

Figure  \ref{Fig:Veff2} shows the behaviour of the effective potentials for some small spherical ($\kappa=1$) black holes with $\Lambda=0$ and $d=5,6$. For the $d=5$ example, the plot for the scalar modes reveals that $V_{eff}<0$ near the horizon. This gives rise to the instability discovered in Ref.~\cite{scalarperts}. However, from our perspective this is much worse than an instability: $c_S<0$ implies that the theory is not hyperbolic near the horizon. Similarly, for the $d=6$ example the tensors have $V_{eff}<0$ near the horizon, which gives the instability of Ref.~\cite{tensorperts,tensorperts2}. But again, this is much worse than an instability, it is a failure of hyperbolicity of the theory.

Finally, in Fig.~\ref{Fig:Veffads} we show an asymptotically AdS example (we take the asymptotic AdS radius to be $\ell=1$ 
where $\Lambda\equiv -(d-1)(d-2)/2\ell^2$%
) with a planar horizon ($\kappa=0$). In this case we have $c_S<c_V<1<c_T$ so the tensors are always superluminal. Near the horizon we have $c_S<0$ so there is a violation of hyperbolicity. If one assumes the existence of a dual CFT then causality of the CFT imposes restrictions on the parameters of the bulk theory. This excludes examples such as the one just discussed, as well as other examples in which hyperbolicity is not violated but the bulk superluminality leads to unacceptable boundary superluminality \cite{brigante,brigante2,hofman,Camanho:2009hu,Camanho:2010ru}. 

\subsection{Conditions for hyperbolicity violation in small black holes}

The signatures of the effective metrics are determined by the sign of $c_A$.  At large $r$, $c_A\rightarrow1$, so all the effective metrics approach the physical metric asymptotically and hence have Lorentzian signature.  As we have seen, this may not be the case closer to the horizon.  We would like to determine situations where such a sign change occurs.  For simplicity, let us focus on small black holes (assuming solutions exist for arbitrarily small horizon radius $r_0$) and on non-planar black holes ($\kappa\neq0$).  

Near the horizon, the effective potential goes as 
\be
V_{eff}=\frac{c_A(r_0)f'(r_0)}{2r_0^2}(r-r_0)\;,
\ee
with $f'(r_0)>0$.  Therefore, the sign of $c_A(r_0)$ determines the signature just outside the horizon.  The horizon is defined by $f(r_0)=0$, so $\psi(r_0)=\kappa/r_0^2$.   Let us evaluate the functions $A$ and $B$, given by \eqref{ABeq}, at the horizon and expand in powers of $r_0$, assuming solutions exist for arbitrarily small $r_0$.  We get
\be
A(r_0)=1-\frac{d-1}{d-3}\frac{P-1}{P}+\mathcal O(r_0^{-2})\;,\qquad B(r_0)=\frac{(d-1)^2}{(d-3)(d-4)}\frac{(P-1)(P-2)}{A(r_0)P^2}+\mathcal O(r_0^{-2})\;,
\ee
where $P$ is the order of the polynomial $W[\psi]$ in \eqref{algrel}.  Note that the lowest order term in $A$ vanishes if $d=1+2P$.  For now, let us assume that $d\neq 1+2P$ so the lowest order term is non-vanishing.  Then we can just put these expressions into \eqref{ceq} to get
\begin{subequations}
\begin{align}
c_T(r_0)&=\frac{d-1-3P}{(d-4)P}+\mathcal O(r_0^2)\\
c_V(r_0)&=\frac{d-1-2P}{(d-3)P}+\mathcal O(r_0^2)\\
c_S(r_0)&=\frac{d-1-P}{(d-2)P}+\mathcal O(r_0^2)\;.
\end{align}
\end{subequations}
Since here $d>1+2P$, there is no sign change for the scalars or vectors.  The tensors, on the other hand, can be negative
if $d=6$ or $d\geq 8$.

Now let us assume $d=1+2P$.  Then the sign of $c_V(r_0)$ can only be determined after expanding to higher powers in $r_0$. $c_T$ and $c_S$ can still be determined to lowest order in $r_0$, but will be different from before due to a number of cancellations. They are given by
\begin{subequations}
\begin{align}
c_T(r_0)&=\frac{2(P-Q)-1}{2P-3}+\mathcal O(r_0^2)
\\
c_V(r_0)&=
-\frac{(P-Q)(P-Q-1)}{P(P-1)}\frac{\tilde\alpha_Q}{\tilde\alpha_P}\left(\frac{r_0^2}{\kappa}\right)^{P-Q}
+\mathcal O\bigl(r_0^{2(P-Q)-2}\bigr)
\\
\qquad 
c_S(r_0)&=-\frac{2(P-Q)-1}{2P-1}+\mathcal O(r_0^2)
\;,
\end{align}
\end{subequations}
where $Q\geq 1$ is the second largest power in the polynomial $W[\psi]$ in $\eqref{algrel}$,
and we assume $Q<P-1$.  $\tilde\alpha_{P,Q}$ are introduced by $W\simeq \tilde\alpha_P \psi^P + \tilde\alpha_Q \psi^Q$ for convenience.  Note that the scalars are negative and the vectors become negative if $Q<P-1$ and $\tilde\alpha_Q/\tilde\alpha_P>0$.  

When $d=1+2P$ and $Q=P-1$, there are additional cancellations and the values of $c_A$ can be different from above.  We get generically\footnote{These equations are valid for $\tilde\alpha_{P-2}\neq\frac{P-1}{2P}\frac{\tilde\alpha_{P-1}^2}{\tilde\alpha_P}$.  We note however, that as long as $d=1+2P$, $c_T$ and $c_S$ are proportional and differ by a sign to lowest order in $r_0$, so one of them (generically $c_S$) must become negative. }
\begin{subequations}
\begin{align}
c_T(r_0)&= \frac{3}{2P-3}+\mathcal O(r_0^2)\
\\
c_V(r_0)&= \mathcal O\bigl(r_0^{4}\bigr)\;,
\\
c_S(r_0)&= -\frac{3}{2P-1}+\mathcal O(r_0^2)\;.
\end{align}
\end{subequations}
We have thus shown that all small Lovelock black holes with $d<1+3P$ break hyperbolicty, and the ``bad" modes can show up in any (scalar, tensor or vector) sector.  

Let us comment on the planar case with $\kappa=0$ which has previously been studied in detail in \cite{Camanho:2010ru}.  In this case, $c_A$ is given by
\begin{subequations}
\begin{align}
c_T(r_0)&=
\frac{
(d-2)^2(d-3)(d-4)
+2\bigl[
(d-2)(d-6)\tilde\alpha_2\Lambda
-2\tilde\alpha_2^2\Lambda^2
+ 3\tilde\alpha_3\Lambda
\bigr]
}{(d-2)(d-4)\big[(d-2)(d-3)+2\tilde\alpha_2\Lambda\big]}
\\
c_V(r_0)&=
1+\frac{2\tilde\alpha_2\Lambda}{(d-2)(d-3)}
\\
c_S(r_0)&=
\frac{(d-2)^2(d-3)+6\bigl[
(d-2)^2\tilde\alpha_2\Lambda
+2\tilde\alpha_2^2 \Lambda^2
-\tilde\alpha_3 \Lambda
\bigr]}{(d-2)^2\bigl[(d-2)(d-3)+2\tilde\alpha_2\Lambda\bigr]}
\;,
\end{align}
\end{subequations}
where $\tilde\alpha_3 \equiv -64\bigl(\prod_{k=3}^{6}(d-k)\bigr)k_3$.  We note that there are other conditions that place constraints on these parameters.  In the Gauss-Bonnet case $\tilde \alpha_3=0$ for example, not all of these can be negative due to the bound (\ref{bound}) on $\tilde\alpha_2\Lambda$. The scalars are negative in $d=5$ for $-3<\tilde\alpha_2\Lambda<-3/2$, and the tensors are negative in $d=6$ for $-5<\tilde\alpha_2\Lambda<-2\sqrt 6$.

\section{Discussion}

\label{sec:discuss}

In this paper we have discussed Lovelock theories of gravity, focusing on their causal properties, as determined by their characteristic hypersurfaces, and their hyperbolicity. 

We proved that a Killing horizon is a characteristic hypersurface for all gravitational degrees of freedom, generalising a result of Ref. \cite{izumi}. On general grounds, the event horizon of a static black hole is expected to be a Killing horizon so this shows that no signal can escape from the interior of such a black hole. Extending this result to the stationary case would involve proving a Lovelock analogue of Hawking's rigidity theorem \cite{hawking} for stationary black holes. 

We have considered two classes of solutions of Lovelock theories and determined their characteristic hypersurfaces. For the case of Ricci flat type N spacetimes, we found that these are the null hypersurfaces of certain Lorentzian ``effective metrics", with one such metric for each physical degree of freedom of the graviton. The null cones of these effective metrics determine the causal structure of the spacetime, as experienced by gravitational disturbances. We have explained how this result implies that Lovelock theories are hyperbolic in such backgrounds. 

In the case of static black hole solutions, we have again found that characteristics are null hypersurfaces with respect to certain effective metrics. When these are Lorentzian, as is the case for large black holes, the theory is hyperbolic. However, for small black holes, one of the effective metrics can change signature near the horizon. This implies that Lovelock theory is not hyperbolic in such backgrounds. 

Our study of hyperbolicity in black hole backgrounds was restricted to the black hole exterior. It would be interesting to extend this investigation to the black hole interior. 

We should emphasize that the two cases investigated here are atypical. The factorisation of the characteristic polynomial into quadratic factors (and hence the appearance of effective metrics) is a consequence of the special properties of these solutions. For most solutions of Lovelock theories, one would not expect this factorisation. Generically, the normal cone is not a product of quadratic cones, but a higher degree cone \cite{CB}. 

The existence of hyperbolicity-violating solutions raises the question of whether non-hyperbolicity can arise dynamically. Starting from initial data for which the equations are hyperbolic, can time evolution break down because the equations become non-hyperbolic?\footnote{Note that this is different from the possibility (discussed in the Introduction) of time evolution breaking down because a surface of constant time fails to be non-characteristic.} What happens if one tries to form one of these small black holes by gravitational collapse? 

One can consider spherically symmetric gravitational collapse in Lovelock theories coupled to matter (see e.g.\ \cite{Maeda:2005ci,Nozawa:2005uy,Maeda:2006pm,jhingan,Ohashi1}). The violation of hyperbolicity is not apparent in a spherically symmetric reduction of the equations because it is the angular part of the effective metric that undergoes a sign change when hyperbolicity is violated, and hence it is only the angular derivatives that are affected by this sign change.\footnote{Other interesting features of Lovelock theories are also absent in a spherically symmetric reduction of the equations of motion. This is because such theories admit a Birkhoff-like theorem \cite{Zegers}, which implies that there are no gravitational wave degrees of freedom, only matter degrees of freedom. Hence, if matter obeys the dominant energy condition then there will be no superluminal propagation in the reduced theory.}  So it might be possible to find a solution of a Lovelock theory coupled to matter that describes spherically symmetric gravitational collapse to form one of these hyperbolicity-violating black holes. 
However, the interesting question, akin to strong cosmic censorship, is whether this happens {\it generically}, which requires breaking spherical symmetry. If one considers generic non-spherically symmetric perturbations of this collapse solution, then what happens? Does time evolution break down because the equations become non-hyperbolic? Or does the existence of some instability drive the system away from the hyperbolicity-violating solution?
\medskip
\begin{center}
\noindent {\bf Acknowledgments}
\end{center}
\medskip
\noindent
HSR is grateful to Claude Warnick for a useful discussion. This work was supported by the European Research Council grant no. ERC-2011-StG 279363-HiDGR.
N.T.\ was supported by the World Premier International Research Center Initiative (WPI
Initiative), MEXT, Japan, and by JSPS Grant-in-Aid for Scientific Research 25$\cdot$755.
\singlespacing
\bibliography{refs}

\providecommand{\href}[2]{#2}\begingroup\raggedright\begin{thebibliography}{10}

\bibitem{lovelock}
D.~Lovelock, {\it {The Einstein tensor and its generalizations}},  {\em J.\
  Math.\ Phys.} {\bf 12} (1971) 498--501.

\bibitem{aragone}
C.~Aragone, {\it {Stringy Characteristics of Effective Gravity}},  in {\em
  SILARG 6}, pp.~60--69, 1987.

\bibitem{CB}
Y.~Choquet-Bruhat, {\it {The Cauchy Problem for Stringy Gravity}},  {\em J.\
  Math.\ Phys.} {\bf 29} (1988) 1891--1895.

\bibitem{brigante}
M.~Brigante, H.~Liu, R.~C. Myers, S.~Shenker, and S.~Yaida, {\it {Viscosity
  Bound Violation in Higher Derivative Gravity}},  {\em Phys.\ Rev.} {\bf D77}
  (2008) 126006, [\href{http://xxx.lanl.gov/abs/0712.0805}{{\tt
  arXiv:0712.0805}}].

\bibitem{brigante2}
M.~Brigante, H.~Liu, R.~C. Myers, S.~Shenker, and S.~Yaida, {\it {The Viscosity
  Bound and Causality Violation}},  {\em Phys.\ Rev.\ Lett.} {\bf 100} (2008)
  191601, [\href{http://xxx.lanl.gov/abs/0802.3318}{{\tt arXiv:0802.3318}}].

\bibitem{hofman}
D.~M. Hofman, {\it {Higher Derivative Gravity, Causality and Positivity of
  Energy in a UV complete QFT}},  {\em Nucl.\ Phys.} {\bf B823} (2009)
  174--194, [\href{http://xxx.lanl.gov/abs/0907.1625}{{\tt arXiv:0907.1625}}].

\bibitem{Camanho:2009hu}
X.~O. Camanho and J.~D. Edelstein, {\it {Causality in AdS/CFT and Lovelock
  theory}},  {\em JHEP} {\bf 1006} (2010) 099,
  [\href{http://xxx.lanl.gov/abs/0912.1944}{{\tt arXiv:0912.1944}}].

\bibitem{Camanho:2010ru}
X.~O. Camanho, J.~D. Edelstein, and M.~F. Paulos, {\it {Lovelock theories,
  holography and the fate of the viscosity bound}},  {\em JHEP} {\bf 1105}
  (2011) 127, [\href{http://xxx.lanl.gov/abs/1010.1682}{{\tt
  arXiv:1010.1682}}].

\bibitem{izumi}
K.~Izumi, {\it {Causal Structures in Gauss-Bonnet gravity}},
  \href{http://xxx.lanl.gov/abs/1406.0677}{{\tt arXiv:1406.0677}}.

\bibitem{Deruelle:1989fj}
N.~Deruelle and L.~Farina-Busto, {\it {The Lovelock Gravitational Field
  Equations in Cosmology}},  {\em Phys.\ Rev.} {\bf D41} (1990) 3696.

\bibitem{teitelboim}
C.~Teitelboim and J.~Zanelli, {\it {Dimensionally continued topological
  gravitation theory in Hamiltonian form}},  {\em Class.\ Quant.\ Grav.} {\bf
  4} (1987) L125.

\bibitem{cmpp}
A.~Coley, R.~Milson, V.~Pravda, and A.~Pravdova, {\it {Classification of the
  Weyl tensor in higher dimensions}},  {\em Class.\ Quant.\ Grav.} {\bf 21}
  (2004) L35--L42, [\href{http://xxx.lanl.gov/abs/gr-qc/0401008}{{\tt
  gr-qc/0401008}}].

\bibitem{pravda}
T.~Malek and V.~Pravda, {\it {Type III and N solutions to quadratic gravity}},
  {\em Phys.\ Rev.} {\bf D84} (2011) 024047,
  [\href{http://xxx.lanl.gov/abs/1106.0331}{{\tt arXiv:1106.0331}}].

\bibitem{boulware}
D.~G. Boulware and S.~Deser, {\it {String Generated Gravity Models}},  {\em
  Phys.\ Rev.\ Lett.} {\bf 55} (1985) 2656.

\bibitem{wheeler}
J.~T. Wheeler, {\it {Symmetric Solutions to the Maximally {Gauss-Bonnet}
  Extended Einstein Equations}},  {\em Nucl.\ Phys.} {\bf B273} (1986) 732.

\bibitem{cai}
R.-G. Cai, {\it {Gauss-Bonnet black holes in AdS spaces}},  {\em Phys.\ Rev.}
  {\bf D65} (2002) 084014, [\href{http://xxx.lanl.gov/abs/hep-th/0109133}{{\tt
  hep-th/0109133}}].

\bibitem{cai2}
R.-G. Cai, {\it {A Note on thermodynamics of black holes in Lovelock gravity}},
   {\em Phys.\ Lett.} {\bf B582} (2004) 237--242,
  [\href{http://xxx.lanl.gov/abs/hep-th/0311240}{{\tt hep-th/0311240}}].

\bibitem{tensorperts}
G.~Dotti and R.~J. Gleiser, {\it {Gravitational instability of
  Einstein-Gauss-Bonnet black holes under tensor mode perturbations}},  {\em
  Class.\ Quant.\ Grav.} {\bf 22} (2005) L1,
  [\href{http://xxx.lanl.gov/abs/gr-qc/0409005}{{\tt gr-qc/0409005}}].

\bibitem{tensorperts2}
G.~Dotti and R.~J. Gleiser, {\it {Linear stability of Einstein-Gauss-Bonnet
  static spacetimes. Part I. Tensor perturbations}},  {\em Phys.\ Rev.} {\bf
  D72} (2005) 044018, [\href{http://xxx.lanl.gov/abs/gr-qc/0503117}{{\tt
  gr-qc/0503117}}].

\bibitem{scalarperts}
R.~J. Gleiser and G.~Dotti, {\it {Linear stability of Einstein-Gauss-Bonnet
  static spacetimes. Part II: Vector and scalar perturbations}},  {\em Phys.\
  Rev.} {\bf D72} (2005) 124002,
  [\href{http://xxx.lanl.gov/abs/gr-qc/0510069}{{\tt gr-qc/0510069}}].

\bibitem{Konoplya:2008ix}
R.~Konoplya and A.~Zhidenko, {\it {(In)stability of D-dimensional black holes
  in Gauss-Bonnet theory}},  {\em Phys.\ Rev.} {\bf D77} (2008) 104004,
  [\href{http://xxx.lanl.gov/abs/0802.0267}{{\tt arXiv:0802.0267}}].

\bibitem{takahashi}
T.~Takahashi and J.~Soda, {\it {Stability of Lovelock Black Holes under Tensor
  Perturbations}},  {\em Phys.\ Rev.} {\bf D79} (2009) 104025,
  [\href{http://xxx.lanl.gov/abs/0902.2921}{{\tt arXiv:0902.2921}}].

\bibitem{takahashi2}
T.~Takahashi and J.~Soda, {\it {Instability of Small Lovelock Black Holes in
  Even-dimensions}},  {\em Phys.\ Rev.} {\bf D80} (2009) 104021,
  [\href{http://xxx.lanl.gov/abs/0907.0556}{{\tt arXiv:0907.0556}}].

\bibitem{Takahashi:2010ye}
T.~Takahashi and J.~Soda, {\it {Master Equations for Gravitational
  Perturbations of Static Lovelock Black Holes in Higher Dimensions}},  {\em
  Prog.\ Theor.\ Phys.} {\bf 124} (2010) 911--924,
  [\href{http://xxx.lanl.gov/abs/1008.1385}{{\tt arXiv:1008.1385}}].

\bibitem{Takahashi:2010gz}
T.~Takahashi and J.~Soda, {\it {Catastrophic Instability of Small Lovelock
  Black Holes}},  {\em Prog.\ Theor.\ Phys.} {\bf 124} (2010) 711--729,
  [\href{http://xxx.lanl.gov/abs/1008.1618}{{\tt arXiv:1008.1618}}].

\bibitem{courant}
R.~Courant and D.~Hilbert, {\em Methods of Mathematical physics}, vol.~2.
\newblock Wiley, 1962.

\bibitem{hawking}
S.~Hawking and G.~Ellis, {\em The large scale structure of space-time}.
\newblock Cambridge University Press, 1973.

\bibitem{wald}
R.~Wald, {\em General Relativity}.
\newblock Oxford University Press, 1984.

\bibitem{GHP}
M.~Durkee, V.~Pravda, A.~Pravdova, and H.~S. Reall, {\it {Generalization of the
  Geroch-Held-Penrose formalism to higher dimensions}},  {\em Class.\ Quant.\
  Grav.} {\bf 27} (2010) 215010, [\href{http://xxx.lanl.gov/abs/1002.4826}{{\tt
  arXiv:1002.4826}}].

\bibitem{deser}
S.~Deser and A.~Ryzhov, {\it {Curvature invariants of static spherically
  symmetric geometries}},  {\em Class.\ Quant.\ Grav.} {\bf 22} (2005)
  3315--3324, [\href{http://xxx.lanl.gov/abs/gr-qc/0505039}{{\tt
  gr-qc/0505039}}].

\bibitem{Maeda:2005ci}
H.~Maeda, {\it {Effects of Gauss-Bonnet terms on final fate of gravitational
  collapse}},  {\em Class.Quant.Grav.} {\bf 23} (2006) 2155,
  [\href{http://xxx.lanl.gov/abs/gr-qc/0504028}{{\tt gr-qc/0504028}}].

\bibitem{Nozawa:2005uy}
M.~Nozawa and H.~Maeda, {\it {Effects of lovelock terms on the final fate of
  gravitational collapse: Analysis in dimensionally continued gravity}},  {\em
  Class.Quant.Grav.} {\bf 23} (2006) 1779--1800,
  [\href{http://xxx.lanl.gov/abs/gr-qc/0510070}{{\tt gr-qc/0510070}}].

\bibitem{Maeda:2006pm}
H.~Maeda, {\it {Final fate of spherically symmetric gravitational collapse of a
  dust cloud in Einstein-Gauss-Bonnet gravity}},  {\em Phys.Rev.} {\bf D73}
  (2006) 104004, [\href{http://xxx.lanl.gov/abs/gr-qc/0602109}{{\tt
  gr-qc/0602109}}].

\bibitem{jhingan}
S.~Jhingan and S.~G. Ghosh, {\it Inhomogeneous dust collapse in d-5
  einstein-gauss-bonnet gravity},  {\em Phys.Rev.} {\bf D81} (2010) 024010.

\bibitem{Ohashi1}
S.~Ohashi, T.~Shiromizu, and S.~Jhingan, {\it {Spherical collapse of
  inhomogeneous dust cloud in the Lovelock theory}},  {\em Phys.\ Rev.} {\bf
  D84} (2011) 024021, [\href{http://xxx.lanl.gov/abs/1103.3826}{{\tt
  arXiv:1103.3826}}].

\bibitem{Zegers}
R.~Zegers, {\it {Birkhoff's theorem in Lovelock gravity}},  {\em J.\ Math.\
  Phys.} {\bf 46} (2005) 072502,
  [\href{http://xxx.lanl.gov/abs/gr-qc/0505016}{{\tt gr-qc/0505016}}].

\end{thebibliography}\endgroup
\bibliographystyle{JHEP}
\end{document}